\documentclass[hidelinks]{llncs}

\usepackage[toc,shortcuts,nonumberlist,smallcaps,numberedsection=autolabel]{glossaries} 

\usepackage{amsthm}
\usepackage{graphicx}
\usepackage{textcomp}
\usepackage{bm}
\usepackage{amssymb}
\usepackage{amsmath}
\usepackage{mathtools}
\usepackage{stmaryrd}
\usepackage[linesnumbered, ruled, noend]{algorithm2e} 
\usepackage{fixltx2e}
\usepackage{tikz} 
\usetikzlibrary{shapes,trees,arrows,backgrounds,automata,positioning,fit,decorations.pathreplacing,calc,shapes.multipart,chains,petri,backgrounds,calc}
\usepackage{booktabs} 
\usepackage{tabularx} 
\usepackage{longtable} 
\usepackage{csvsimple} 
\usepackage{rotating} 
\usepackage{listings} 
\usepackage{hyperref}
\usepackage{caption}
\usepackage{enumitem} 
\usepackage{cleveref}
\usepackage{caption}
\usepackage{wrapfig}
\usepackage{array} 
\usepackage{varwidth} 
\usepackage{cases}
\usepackage{upgreek}
\usepackage{varwidth} 
\usepackage{empheq}
\usepackage[nocompress]{cite}
\usepackage{todonotes}
\usepackage{algorithmicx}
\usepackage{chngcntr}
\usepackage{pbox}
\usepackage{multirow}
\usepackage{subfig}

\newacronym{lps}{lps}{Linear Process Specification}
\newacronym{pts}{pts}{Partitioned Transition System}
\newacronym{rdm}{rdm}{Read Dependency Matrix}
\newacronym{wdm}{wdm}{Must-write Dependency Matrix}
\newacronym{mdm}{mdm}{May-write Dependency Matrix}
\newacronym{ldd}{ldd}{List Decision Diagram}
\newacronym{dag}{dag}{Directed A-cyclic Graph}
\newacronym{por}{por}{Partial Order Reduction}
\newacronym{ts}{ts}{Transition System}   
\newacronym{mdd}{mdd}{Multi-way Decision Diagram}
\newacronym{bdd}{bdd}{Binary Decision Diagram}
\newacronym{dm}{dm}{Dependency Matrix}
\newacronym{sus}{sus}{State Update Specification}
\newacronym{nes}{nes}{Normalized Event Span}
\newacronym{wes}{wes}{Weighted Event Span}
\makeglossaries

\tikzstyle{bdd-leaf}=[circle,draw,solid,thick,minimum size=6mm, inner sep=0mm]
\tikzstyle{ldd-var-two}=[rectangle,draw, solid,minimum size=6mm, inner sep=0mm, rectangle split, rectangle split parts=2, rectangle split horizontal, align=center, text width=.6cm]
\tikzstyle{ldd-var}=[rectangle,draw, solid,minimum size=6mm, inner sep=0mm]
\tikzstyle{ldd}=[node distance=2cm, every edge/.style={-stealth, draw}]
\tikzstyle{normal}=[out=270, in=90]
\tikzstyle{false}=[out=0, in=90, max distance=1.9cm]


\lstdefinestyle{multiline}{
basicstyle=\scriptsize\sffamily,
numbers=left,
numberstyle=\tiny,
frame=tb,
columns=fullflexible,
showstringspaces=false,
escapeinside=\`\`
}

\let\bbordermatrix\bordermatrix
\patchcmd{\bbordermatrix}{8.75}{4.75}{}{}
\patchcmd{\bbordermatrix}{\left(}{\left[}{}{}  
\patchcmd{\bbordermatrix}{\right)}{\right]}{}{}

\let\abordermatrix\bordermatrix
\patchcmd{\abordermatrix}{8.75}{4.75}{}{}
\patchcmd{\abordermatrix}{\left(}{\left\langle}{}{}
\patchcmd{\abordermatrix}{\right)}{\right\rangle}{}{}
\usepackage{etoolbox}

\AtBeginDocument{%
  \mathchardef\mathcomma\mathcode`\,
  \mathcode`\,="8000 
}
{\catcode`,=\active
  \gdef,{\mathcomma\discretionary{}{}{}}
}

\newcolumntype{M}{>{\begin{varwidth}{1.5cm}}l<{\end{varwidth}}} 
\newenvironment{tagcases}[1][]
  {\empheq[left={#1\empheqlbrace}]{alignat=2}}
  {\endempheq}
  
\makeatletter
\newcommand{\raisemath}[1]{\mathpalette{\raisem@th{#1}}}
\newcommand{\raisem@th}[3]{\raisebox{#1}{$#2#3$}}
\makeatother

\counterwithin*{equation}{definition}
\theoremstyle{remark}
\newtheorem{mycase}{Case}
\theoremstyle{remark}
\newtheorem{mysubcase}{Case}
\numberwithin{mysubcase}{mycase}

\makeatletter
\@ifpackagelater{tikz}{2013/12/01}{%
  \def\AtanTwo(#1,#2){atan2({#1},{#2})}%
}{%
  \def\AtanTwo(#1,#2){atan2({#2},{#1})}%
}
\makeatother

\newcommand{\convexpath}[2]{
[   
    create hullnodes/.code={
        \global\edef\namelist{#1}
        \foreach [count=\counter] \nodename in \namelist {
            \global\edef\numberofnodes{\counter}
            \node at (\nodename) [draw=none,name=hullnode\counter] {};
        }
        \node at (hullnode\numberofnodes) [name=hullnode0,draw=none] {};
        \pgfmathtruncatemacro\lastnumber{\numberofnodes+1}
        \node at (hullnode1) [name=hullnode\lastnumber,draw=none] {};
    },
    create hullnodes
]
($(hullnode1)!#2!-90:(hullnode0)$)
\foreach [
    evaluate=\currentnode as \previousnode using \currentnode-1,
    evaluate=\currentnode as \nextnode using \currentnode+1
    ] \currentnode in {1,...,\numberofnodes} {
-- ($(hullnode\currentnode)!#2!-90:(hullnode\previousnode)$)
  let \p1 = ($(hullnode\currentnode)!#2!-90:(hullnode\previousnode) - (hullnode\currentnode)$),
    \n1 = {\AtanTwo(\y1,\x1)},
    \p2 = ($(hullnode\currentnode)!#2!90:(hullnode\nextnode) - (hullnode\currentnode)$),
    \n2 = {\AtanTwo(\y2,\x2)},
    \n{delta} = {-Mod(\n1-\n2,360)}
  in 
    {arc [start angle=\n1, delta angle=\n{delta}, radius=#2]}
}
-- cycle
}

\providecommand{\tuple}[1]{\left( #1 \right)}
\providecommand{\set}[1]{\left\lbrace #1 \right\rbrace}
\providecommand{\set}[1]{\left\lbrace #1 \right\rbrace}
\providecommand{\setcompl}[1]{\overline{ #1 }}

\providecommand{\sizeof}[1]{\left\vert{#1}\right\vert}
\providecommand{\scope}[1]{\left( #1 \right)}

\newcommand{\powerset}[1]{2^{#1}}
\providecommand{\abs}[1]{\lvert#1\rvert}

\newcommand{\suchthat}{\mathrel{.}}

\newcommand{\pins}{P\textsc{ins}\xspace}

\newcommand{\ltsmin}{\textsc{LTSmin}\xspace}

\newcommand{\natnumset}{\mathbb{N}}
\providecommand{\natnumsetleq}[1]{\natnumset_{\mathsmaller{\leq {#1} }}}
\providecommand{\natnumsetlt}[1]{\natnumset_{\mathsmaller{< {#1} }}}

\newcommand{\mcrl}{m\textsc{crl}2\xspace}

\newcommand{\promela}{\textsc{Promela}\xspace}   

\newcommand{\dve}{\textsc{Dve}\xspace}   

\newcommand{\uppaal}{\textsc{Uppaal}\xspace}   

\newcommand{\divine}{\textsc{d}i\textsc{v}in\textsc{e}\xspace}

\newcommand{\nusmv}{\textsc{n}u\textsc{smv}\xspace}

\newcommand{\smart}{\textsc{smart}\xspace}

\newcommand{\pnml}{\textsc{pnml}\xspace}

\newcommand{\boost}{Boost\xspace}

\newcommand{\viennacl}{ViennaCL\xspace}

\newcommand{\marcie}{\textsc{marcie}\xspace}

\newcommand{\force}{\textsc{force}\xspace}

\let\emptyset\varnothing
\providecommand{\order}[2]{\ensuremath{\mathop{O_{ #1 }^{ #2}}}\xspace}
\providecommand{\orderfn}[2]{\ensuremath{\mathop{p_{ #1 }}( #2 )}}

\providecommand{\posdef}[1]{\ensuremath{p_{\mathsmaller{ #1 }}}\xspace}
\providecommand{\pos}[2]{\ensuremath{\mathop{\posdef{ #1 }}( #2 )}\xspace}
\newcommand{\permdef}{\ensuremath{\pi}\xspace}
\providecommand{\perm}[1]{\ensuremath{{\mathop{\permdef}}( #1 )}\xspace}

\newcommand{\mtrx}{\ensuremath{\bm{A}}\xspace}
\newcommand{\amtrx}{\ensuremath{\hat{\mtrx}}\xspace}

\newcommand{\numrows}{\ensuremath{\textsc{m}}\xspace}
\newcommand{\numcols}{\ensuremath{\textsc{n}}\xspace}
\newcommand{\me}{\ensuremath{\mathrm{a}}\xspace}
\newcommand{\ame}{\ensuremath{\hat{\me}}\xspace}

\providecommand{\zmtrx}[2]{\ensuremath{\bm{0}^\mathsmaller{#1 \times #2}}\xspace}

\newcommand{\graph}{\ensuremath{G}\xspace}
\newcommand{\vertices}{\ensuremath{V}\xspace}
\newcommand{\edges}{\ensuremath{E}\xspace}

\newcommand{\tg}{\ensuremath{\mathop{\graph^T}}\xspace}
\newcommand{\tv}{\ensuremath{\mathop{\vertices^T}}\xspace}
\newcommand{\te}{\ensuremath{\mathop{\edges^T}}\xspace}

\providecommand{\bwdef}[1]{\ensuremath{b_{\mathsmaller{ #1 }}}\xspace}
\providecommand{\bw}[2]{\ensuremath{\mathop{\bwdef{ #1 }}( #2 )}\xspace}

\providecommand{\spndef}[1]{\ensuremath{s_{\mathsmaller{ #1 }}}\xspace}
\providecommand{\spn}[2]{\ensuremath{\mathop{\spndef{ #1 }}( #2 )}\xspace}

\providecommand{\wfdef}[1]{\ensuremath{f_{\mathsmaller{ #1 }}}\xspace}
\providecommand{\wf}[2]{\mathop{\wfdef{ #1 }}( #2 )\xspace}

\newcommand{\writegraph}{\ensuremath{\graph^\mathcal{W}}\xspace}

\providecommand{\neighborsdef}[1]{\ensuremath{n_{\mathsmaller{ #1 }}}\xspace}
\providecommand{\neighbors}[2]{\ensuremath{\mathop{\neighborsdef{#1}}( #2 )}\xspace}

\providecommand{\ordert}[1]{\ensuremath{\mathop{\order{ #1 }{}^{T}}}\xspace}

\newcommand{\pts}{\ensuremath{\mathcal{P}}\xspace}

\newcommand{\p}{\ensuremath{\mathrm{p}}}

\newcommand{\WM}{\ensuremath{\bm{W}}}
\newcommand{\WME}{\ensuremath{\bm{w}}}

\newcommand{\inits}{\ensuremath{\vec{s}^0}}

\newcommand{\writeset}{\ensuremath{\mathcal{W}}}
\newcommand{\vars}{\ensuremath{\mathop{\mathcal{S}^\numcols}}}
\newcommand{\transrels}{\ensuremath{\mathop{\to^\numrows}}}
\newcommand{\states}{\ensuremath{\mathop{S^\numcols}}}

\providecommand{\bigo}[1]{\ensuremath{\mathcal{O}( #1 )}\xspace}
\newcommand{\maxdeg}{\ensuremath{\hat{D}}\xspace}

\begin{document}

\algnewcommand{\algorithmicgoto}{\textbf{go to}}%
\algnewcommand{\Goto}[1]{\algorithmicgoto~\ref{#1}}%
\hyphenation{front-width}

\title{Bandwidth and Wavefront Reduction for Static Variable Ordering in Symbolic Model Checking}

\author{Jeroen Meijer \and Jaco van de Pol}

\institute{Formal Methods and Tools, University of Twente, The Netherlands \\\email{\{j.j.g.meijer, j.c.vandepol\}@utwente.nl}}

\maketitle
\begin{abstract}
We demonstrate the applicability of bandwidth and wavefront reduction algorithms to static variable
ordering. In symbolic model checking event locality plays a major role in time and memory usage.
For example, in Petri nets event locality can be captured by dependency matrices, where nonzero entries indicate whether a 
transition modifies a place. The quality of event locality has been expressed as a metric called (weighted) event span.
The bandwidth of a matrix is a metric indicating the distance of nonzero
elements to the diagonal. Wavefront is a metric indicating the degree of nonzeros on one end of the diagonal of the matrix.
Bandwidth and wavefront are well studied metrics used in sparse matrix solvers.
 
In this work we prove that span is limited by twice the bandwidth of a matrix.
This observation makes bandwidth reduction algorithms useful for obtaining good variable orders.
One major issue we address is that the reduction algorithms can only be applied on symmetric matrices,
while the dependency matrices are asymmetric.
We show that the Sloan algorithm executed on the total graph of the adjacency graph gives the best variable orders.
Practically, we demonstrate that our work allows to call standard sparse matrix operations in \boost
and \viennacl, computing very good static variable orders in milliseconds.
Future work is promising, because a whole new spectrum of more off-the-shelf algorithms,
including metaheuristic ones, become available for variable ordering.
\keywords{bandwidth, profile, wavefront, event span, symbolic model checking, sparse matrix,
event locality, decision diagram, Petri net}
\end{abstract}

\section{Introduction}\label{sec:intro}
Model checking is an approach for finding errors in computer programs by
computing reachable states of a program and evaluating formulas over these set of states. Some type of computer
programs allow efficient storage of its set of reachable states by means of decision diagrams, this technique is known
as symbolic model checking \cite{burch1990symbolic}. Storing sets of states symbolically entails storing
sets of integer vectors as binary formulas in Binary Decision Diagrams (BDDs) \cite{DBLP:journals/tc/Bryant86}
or more recent as Multi-value Decision Diagrams (MDDs) \cite{kam1998multi}. One major issue 
with this approach is the ordering of variables in decision diagrams (DDs) representing the formula. Improving variable
ordering is known to be NP-complete \cite{Bollig:1996:IVO:235652.235653}, thus heuristic \cite{Liu09}
and metaheuristic \cite{DBLP:conf/tacas/SiminiceanuC06} algorithms have been developed to improve static
variable ordering. The static variable ordering approach orders variables before reachability analysis,
while dynamic variable ordering happens during the computation of reachable states. 

Static variable ordering typically exploits the notion of event locality. Events, such as program statements
or transitions in Petri nets are often local, i.e. they modify or read only a few variables or places and
ordering these local variables near each other tends to significantly reduce the memory footprint of
the DDs. A good metric for event locality is called the Weighted Event Span (WES),
by Siminiceaunu et al.\cite{DBLP:conf/tacas/SiminiceanuC06}. The WES metric is used to measure the total normalized
distance between the minimum variable and maximum variable of all events. Furthermore the metric includes 
a moment which signifies the importance of involved events happening in the bottom of the DD.
This is important, because with saturation operations in DDs are cheaper in the bottom, rather than the top.
Every operation in the top of the DD recursively propagates down the DD.

The degree of locality of events can be visualized using matrices. Such an approach is taken in \cite{DBLP:conf/hvc/MeijerKBP14}, where
a dependency matrix has rows as transitions and columns as variables. A nonzero entry indicates that a transition depends on a variable, 
e.g. a transition can either read or write to a variable \cite{DBLP:conf/hvc/MeijerKBP14}. These
dependency matrices tend to be sparse, hinting that traditional sparse matrix algorithms can be applied to these matrices.

A subcategory of sparse matrix algorithms are bandwidth and wavefront reduction algorithms.
One key example of a bandwidth reduction algorithm is by Cuthill and McKee developed in 1969 \cite{Cuthill:1969:RBS:800195.805928}.
The goal of these algorithms is very similar to WES reduction algorithms. The bandwidth measures
the distance of nonzeros from the diagonal of the matrix, while wavefront measures the degree
of nonzeros on one end of the diagonal of the matrix. Bandwidth is related to event span because reducing bandwidth must
also reduce event span, because of the triangle inequality, which states that event span is always smaller
than twice the bandwidth. Since wavefront reduction moves nonzeros to the right of the matrix, wavefront
reduces the moment in the WES metric.

Another popular algorithm in numerical analysis is Sloan's \cite{NME:NME1620281111} algorithm,
which optimizes total bandwidth (also called profile) and wavefront. The graph algorithm has a very low time complexity
$\bigo{\maxdeg \cdot \log \maxdeg \cdot \sizeof{\vertices}}$, where
$\maxdeg$ is the maximum degree, and $V$ the set of vertices. This results in runtimes of mere seconds when
applied to matrices with a million rows and columns -- or transitions and variables. Conveniently,
Sloan's algorithm is freely available in \boost's graph library\footnote{\url{http://www.boost.org/doc/libs/1_59_0/libs/graph/doc/sloan_ordering.htm}}.
Every model checker written in C/C++ or Python can be linked to Boost without much effort. 

While bandwidth and wavefront reduction algorithms have proven themselves during the past decades,
they only work on symmetric matrices. A dependency matrix is asymmetric because clearly, transitions (rows)
and variables (columns) are different objects and there exists no natural total order on the union of both.
Reid et al. \cite{DBLP:journals/siammax/ReidS06} discuss several methods of symmetrizing asymmetric matrices.
With visualizations and experimental data we show that indeed, simply assigning some total order, 
that preserves the partial order on transitions and variables works well for symbolic model checking.


We extensively benchmark the Cuthill McKee, Gibbs Poole Stockmeyer, King and Sloan nodal ordering algorithms
implemented in \boost and \viennacl\footnote{\url{http://viennacl.sourceforge.net}} \cite{Rupp:ViennaCL}.
The benchmark consists of more than 300 Petri net models from the 2015 model checking contest \cite{mcc:2015}.
Both libraries are linked to the \ltsmin model checker. The key feature of \ltsmin
is the \pins architecture, extensively discussed in \cite{eemcs26128}. \ltsmin allows language independent model
checking by exposing the \emph{next-state} function in \pins, which language front-ends, such as \divine, \promela or \uppaal implement.

\begin{figure}
   \vspace{-15pt}
\centering
\resizebox{\textwidth}{!}{%
\begin{tikzpicture}
	\large
	
	\node[draw, minimum size=6mm] (pnml) {\pnml front-end};
	\node[draw, text width=3.8cm, minimum size=10mm, right of = pnml, node distance = 5cm] (reord) {Variable Reordering, Transition Grouping};
	\node[draw, minimum size=6mm, right of = reord, node distance = 5cm] (sym) {Symbolic back-end};
	
	\draw (pnml) edge[thick, -stealth] (reord);
	\draw (reord) edge[thick, -stealth] (sym);

    \node[draw, minimum size=6mm, below of = reord, node distance = 2cm] (tg) {Total graph};
    \node[draw, text width=3cm, minimum size=6mm, left of = tg, node distance = 4cm, align=center] (graph) {Totally ordered adjacency graph};
    \node[draw, text width=2.2cm, minimum size=6mm, left of = graph, node distance = 4.3cm, align=center] (dm) {Dependency matrix};
    \node[draw, minimum size=6mm, right of = tg, node distance = 3.5cm] (perm) {Permutation};
    \node[draw, minimum size=6mm, right of = perm, node distance = 4cm] (dperm) {Split permutation};
    \node[draw, minimum size=6mm, below of = perm, node distance = 1.5cm] (pgraph) {Reordered graph};
    \node[draw, text width=3.5cm, minimum size=6mm, below of = dperm, node distance = 1.75cm, align=center] (pdm) {Reordered \\ dependency matrix};
    
    \path (reord) edge[thick, -stealth] node[above=5pt] {1} (dm)
        (dm) edge[thick, -stealth] node[below] {2}  (graph)
        (graph) edge[thick, -stealth, bend left] node[below] {3}  (perm)
        (graph) edge[thick, -stealth] node[below] {3a} (tg)
        (tg) edge[thick, -stealth] node[below] {3b} (perm)
        (perm) edge[thick, -stealth] node[below] {4} (dperm)
        (dperm) edge[thick, -stealth] node[above=5pt] {5} (reord)
        (perm) edge[thick, -stealth] node[left] {A} (pgraph)
        (dperm) edge[thick, -stealth] node[left] {B} (pdm);
    
\end{tikzpicture}
}
   \vspace{-20pt}
\caption{\pins' extension}
\label{fig:pins:extension}
   \vspace{-20pt}
\end{figure}

The proposed extension to \pins is shown in \autoref{fig:pins:extension}. To perform variable reordering with
nodal ordering algorithms the following steps are added to the \emph{Variable Reordering, Transition Grouping} layer in \pins.
\begin{enumerate}
   \vspace{-5pt}
  \item {Existing feature in \pins. Dependency matrices are stored using a dense matrix format. 
        Associated with the (asymmetric) dependency matrix is a partially ordered adjacency graph.} 
  \item {Assign a total order to the adjacency graph of the dependency matrix using \boost or \viennacl.
         Assigning a total order makes the dependency matrix symmetric.}
  \item {Obtain a permutation using a nodal ordering algorithm.
        \begin{enumerate}
          \item Optionally create a total graph first.
          \item Obtain a permutation using a nodal ordering algorithm.
        \end{enumerate}} 
  \item Create a split permutation for the dependency matrix. I.e. one permutation for the rows and one for the columns.
  \item Make the symbolic back-end use the split permutation for transitions and variables.
  \item[A.] Optionally create the reordered graph to print metrics.
  \item[B.] Optionally create the reordered dependency matrix to print metrics.
   \vspace{-5pt}
\end{enumerate}

\section{Preliminaries}\label{sec:pre}
The bandwidth and wavefront reduction algorithms are designed for undirected ordered graphs. We 
therefore introduce the notion of unordered pairs. We also define orders on sets, which allow us to define ordered graphs.
The biadjacency matrix, which is a matrix representation of a bipartite graph, is important for understanding the dependency matrix.
The two parts of a bipartite graph are the set of transitions and the set of variables.
Additionally we introduce the notion of a total graph. Kaveh \cite{eltit} suggests to create the total graph
of the adjacency graph,
because some algorithms produce even better permutations on the total graph at the expense of additional computation time.

\subsection{Sets}\label{ssec:sets}

The set $\natnumset$ denotes the set of natural numbers excluding $0$, and $\natnumset^0$ including $0$.
The set $\natnumsetleq{\numrows}$ denotes the natural numbers less than or equal to \numrows.
Given a set $S$, an unordered pair of elements from $S$ is denoted $\set{a,b} \in \binom{S}{2} = \set{\set{c,d} \mid c \in S \land d \in S \land c \neq d}$.
We write $S^2 = S \times S$, for the set of ordered pairs.

\begin{definition}[order]
Given a set \vertices, an order on \vertices is a (reflexive, antisymmetric and transitive) relation $\order{}{} \subseteq \vertices^2$.
We write $a \leq b = \tuple{a,b} \in \order{}{}$. If $\forall a,b \in \vertices \colon \tuple{a,b} \in \order{}{} \lor \tuple{b,a} \in O$
then $\order{}{}$ is total, otherwise $\order{}{}$ is partial.
We define function $\posdef{\order{}{}} \colon \vertices \to \natnumsetleq{\sizeof{\vertices}}$,
such that $v \mapsto \sizeof{\set{u \mid \tuple{u,v} \in \order{}{}}}$ gives the position of an element in an order.
If $\order{}{}$ is clear from the context, we write $\orderfn{}{v}$ as shorthand for $\pos{\order{}{}}{v}$.
Given a set $U \subseteq \vertices$ we write $\order{U}{} = \order{}{} \cap \mathop{U^2}$.
Given a permutation $\permdef \colon \vertices \to \vertices$, a permuted order $\order{}{\permdef}$ on \vertices under \permdef is
$a \leq_{\permdef} b = \tuple{a,b} \in \order{}{\permdef} \iff \perm{a} \leq \perm{b} = \tuple{\perm{a},\perm{b}} \in \order{}{}$.
To avoid ambiguity, we note $\order{U}{\permdef} = \scope{\order{}{}^\permdef}_U$.
\end{definition}

\subsection{Graphs}\label{ssec:graphs}

\begin{definition}[undirected ordered graph]
An \emph{undirected ordered graph} is a triple $\graph = \tuple{\vertices, \edges, \order{}{}}$,
where \vertices is a set of vertices, $\set{a,b} \in \edges \subseteq \binom{V}{2}$ is a set of undirected edges and
$\order{}{} \subseteq \vertices^2$ defines an order on \vertices.
A bipartite graph is denoted $\graph = \tuple{P \cup \setcompl{P}, \edges, \order{}{}}$, with no edges between nodes in $P$ nor $\setcompl{P}$.
The function $\neighborsdef{\graph} \colon \vertices \to \powerset{\vertices}$,
such that $u \mapsto \set{v \mid \set{u,v} \in \edges} \setminus \set{u}$ gives all the neighbors of a vertex.
The degree of a vertex $v$ is $\sizeof{\neighbors{}{v}}$ (in this paper all graphs are simple graphs).
\end{definition}

\begin{definition}[adjacency matrix]\label{def:matrix}
Given a graph $\graph = \tuple{\vertices, \edges, \order{}{}}$ the \emph{adjacency matrix} of \graph is
a function $\natnumsetleq{\sizeof{\vertices}} \times \natnumsetleq{\sizeof{\vertices}} \to \set{0,1}$, also denoted
\[\amtrx = 
\begin{bmatrix} 
\ame_{11} & \ame_{12} & \cdots & \ame_{1\sizeof{\vertices}} \\
\ame_{21} & \ame_{22} & \cdots & \ame_{2\sizeof{\vertices}} \\ 
\vdots & \vdots & \ddots & \vdots \\ 
\ame_{\sizeof{\vertices}1} & \ame_{\sizeof{\vertices}2} & \cdots & \ame_{\sizeof{\vertices}\sizeof{\vertices}}
\end{bmatrix}
\in \set{0,1}^{\sizeof{\vertices} \times \sizeof{\vertices}},\text{ such that}\]

\begin{tagcases}[\tuple{i,j} \mapsto]
0 & \text{ if } \set{u,v} \not \in \edges \land \orderfn{}{u} = i \land \orderfn{}{v} = j, \notag \\
1 & \text{ if } \set{u,v} \in \edges \land \orderfn{}{u} = i \land \orderfn{}{v} = j. \notag
\end{tagcases}
If \graph is bipartite with parts $R \cup C = \vertices$ and $\forall r \in R\suchthat\forall c \in C \colon r < c$
the adjacency matrix is symmetric, of the form
$
\amtrx = \begin{bmatrix}
\zmtrx{\sizeof{R}}{\sizeof{R}} & \mtrx \\
\mtrx^T & \zmtrx{\sizeof{C}}{\sizeof{C}}
\end{bmatrix}
$
and \mtrx is called the biadjacency matrix. The dependency matrix is this biadjacency matrix.
The element at $\tuple{i,j}$ of an $M \times N$ matrix is on the diagonal iff $i = j$.
\end{definition} 

\begin{definition}[total graph]
Given an undirected ordered graph $\graph = \tuple{\vertices, \edges, \order{}{}}$, a total graph of \graph is 
$\tg = \tuple{\tv, \te, \ordert{}}$, where $\tv = \vertices \cup \edges$ is a set of vertices and
$\te = \edges \cup \{\set{a, \set{a, b}} \mid \set{a, b} \in \edges\} \cup
\{\{\set{a, c}, \set{c, b}\} \mid \{\set{a, c}, \set{c, b}\} \subseteq \edges\} \subseteq \binom{\tv}{2}$ is the set of edges, and
$\ordert{} \subseteq \order{}{} \cup \tv^2$,
i.e. we add all possible vertex-edge edges, edge-vertex edges and edge-edge edges and assign some order
to \tv. 

\noindent\emph{Throughout this work we use the order on \tg where all edges from $\graph$ are larger than all vertices from $\graph$ and all edges are
mutually ordered in lexicographic order.}
\end{definition}

\section{Background}\label{sec:background}
\begin{wrapfigure}{r}{0.4\textwidth}
\vspace{-25pt}
\centering
\resizebox{.41\textwidth}{103pt}{
\begin{tikzpicture}[node distance=1.7cm,>=stealth',bend angle=45,auto]
    
    \node[label={[label distance=.1cm]180:$p_4$}, place,tokens=1] (P1) {};    
    \node[node distance=3cm, label={[label distance=.1cm]180:$p_2$}, place,tokens=0, above left of=P1] (P2) {};
    \node[node distance=3cm, label={[label distance=.1cm]0:$p_5$}, place,tokens=0, above right of=P1] (P4) {};
    \node[node distance=3cm, label={[label distance=.1cm]180:$p_3$}, place,tokens=0, below left of=P1] (P3) {};
    \node[node distance=3cm, label={[label distance=.1cm]0:$p_1$}, place,tokens=0, below right of=P1] (P5) {};
    \node [transition] (T1) [node distance=1.5cm, above of=P1] {$t_1$} 
      edge [pre]                  (P1)
      edge [post]                 (P2)
      edge [post]                 (P4);
    \node [transition] (T3) [left of=P1] {$t_3$}
      edge [pre]                  (P3)
      edge [post]                 (P2);
    \node [transition, node distance=.8cm] (T2) [left of=T3] {$t_2$}
      edge [pre]                  (P2)
      edge [post]                 (P3);
    \node [transition] (T4) [right of=P1] {$t_4$}
      edge [pre]                  (P4)
      edge [post]                 (P5);
    \node [transition, node distance=.8cm] (T5) [right of=T4] {$t_5$}
      edge [pre]                  (P5)
      edge [post]                 (P4);
    \node [transition] (T6) [node distance=1.5cm, below of=P1] {$t_6$}
      edge [pre]                  (P5)
      edge [pre]                  (P3)
      edge [post]                 (P1);
           
\end{tikzpicture}
}
   \vspace{-20pt}
\captionof{figure}{Example 1-safe Petri net.}
\label{fig:petrinet}
   \vspace{-20pt}
\end{wrapfigure}
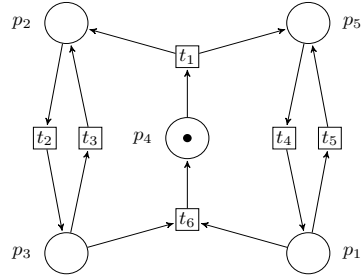
Model checking involves answering questions such as whether or not a system can enter a particular state.
Consider \Cref{fig:petrinet}, which is an example of a (1-safe) Petri net. A Petri net is a bipartite graph
where its vertices are places and transitions. Places can contain a positive number of tokens. The graph's
edges are called arcs.
An outgoing arc means that tokens will be consumed and incoming arc means that tokens will be produced.
In \Cref{fig:petrinet}, after transition $t_1$ fires, $p_4$ will have no token, while both $p_2$ and $p_5$ get one token.
A reachability question is whether or not $p_1$ will eventually have a token, which it will, after firing $t_1$ followed by $t_4$.
A Petri net can model many kinds of systems or protocols such as Lamport's mutual exclusion algorithm,
or they can even model biological processes. 

\begin{figure}%
   \vspace{-25pt}
\centering
\subfloat[Initial order]{%
{\includegraphics[height=150pt]{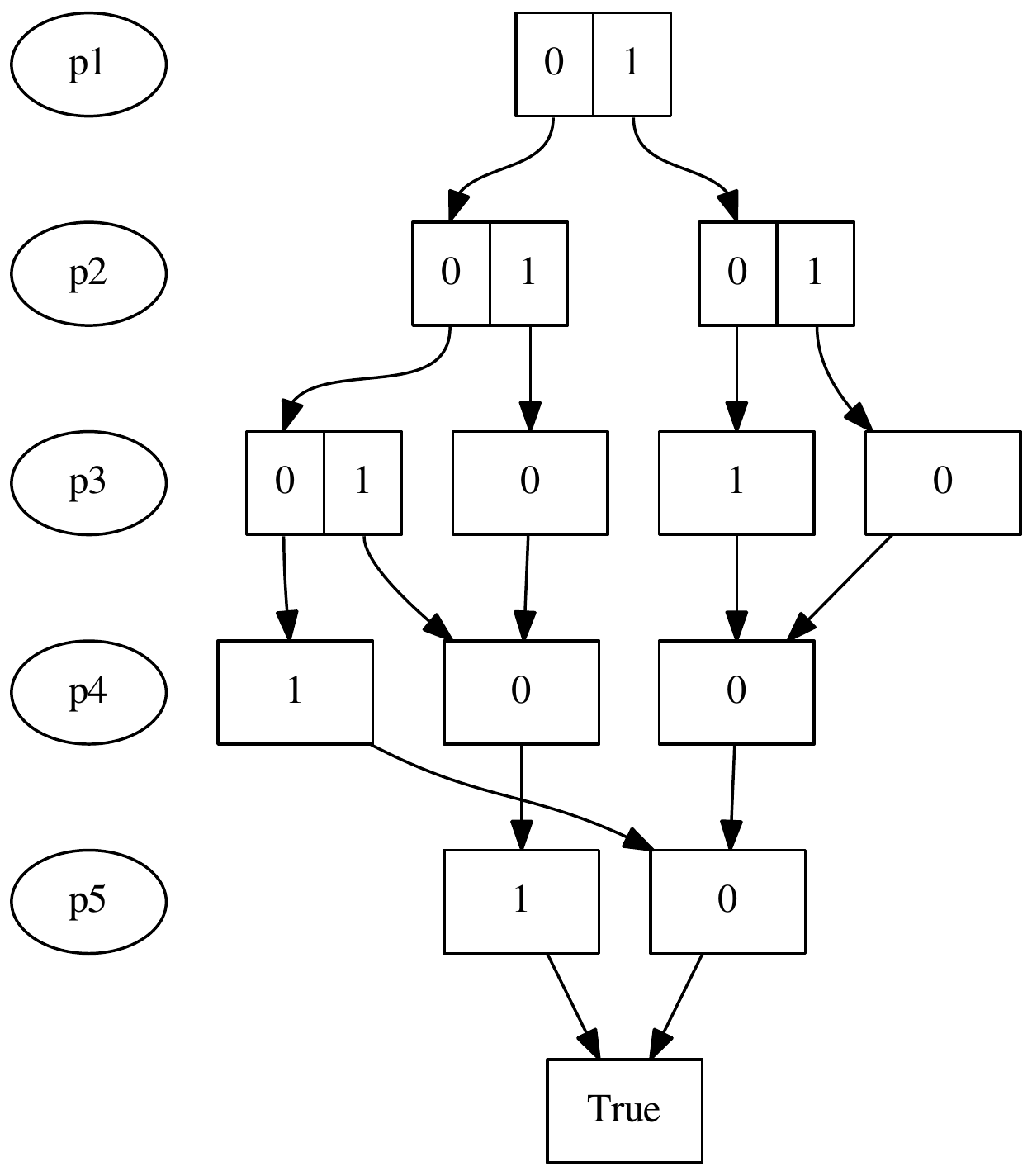}\label{fig:ldd:init} }}%
\qquad
\subfloat[Cuthill McKee]{%
{\includegraphics[height=150pt]{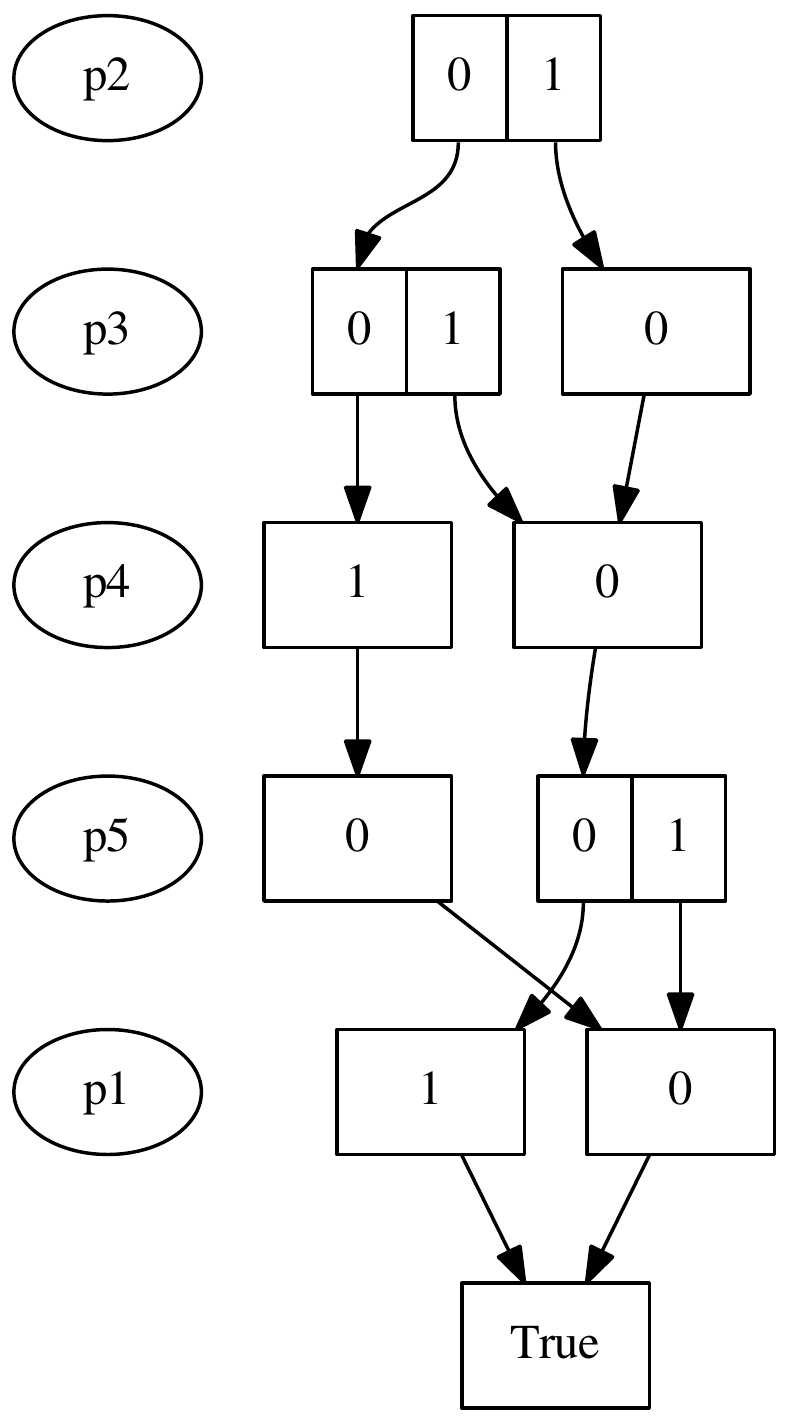}\label{fig:ldd:cm} }}%
   \vspace{-5pt}
\caption{Reachable states}%
\label{fig:ldd}%
   \vspace{-15pt}
\end{figure}

The key issue with storing all reachable states in decision diagrams is the ordering of variables, i.e.
the order of places $p_1,p_2,p_3,p_4,p_5$. \Cref{fig:ldd} shows particular decision diagrams,
namely List Decision Diagrams (LDDs) \cite{dijktsylvan}, representing the five reachable states of the Petri net, but
with different variable orders. Every path from the top left node to the \texttt{true} node represents
a reachable state. The value in a node indicates the number of tokens.
One can see that \Cref{fig:ldd:cm}, whose variable order is computed
using Cuthill McKee, has fewer nodes than \Cref{fig:ldd:init} with the default alphanumeric variable order.
Thus to store the decision diagram in \Cref{fig:ldd:cm} requires less memory. 

\subsection{Symbolic Reachability Analysis}\label{ssec:symmod}

Symbolic reachability analysis is a strategy that repeatedly applies the transition relation until a fixed point is reached.
For efficient symbolic reachability analysis the transition relation can be partitioned \cite{burch1991symbolic}.
The relation can be partitioned conjunctively, for synchronous systems, or disjunctively for
asynchronous systems. The latter approach is used for the Petri net in \autoref{fig:petrinet}.
The following notion of a Partitioned Transition System (PTS) can be used to represent any asynchronous system.
Furthermore, PTSs allow us to describe event locality in terms of write dependencies and read dependencies. 

\begin{definition}[\acrlong{pts}]
\label{def:pts}
A \emph{\acrlong{pts}} (PTS) is a structure 
$\pts = \tuple{\states, \transrels, \inits}$, where
\begin{itemize}[topsep=2pt,parsep=0pt,partopsep=2pt]
\item $\states = \prod_{i=1}^\numcols S^\numcols_i$ is the set of states 
$\vec{s} \in \states$, which are vectors of \numcols values,
\item $\transrels = \bigcup_{i=1}^\numrows {\to_i}$ 
is the transition relation,
which is a union of the \numrows transition groups 
${\to_i} \subseteq \states^2$ (for $1 \leq i \leq \numrows$), and
\item $\inits = \tuple{s_1^0, \dotsc, s_\numcols^0} \in \states$ 
is the initial state.
\end{itemize}
We write $\vec{s} \to_i \vec{t}$ when $\tuple{\vec{s}, \vec{t}} \in {\to_i}$ 
for $1 \leq i \leq \numrows$,
$\vec{s} \to_\pts \vec{t}$ when
$\tuple{\vec{s}, \vec{t}} \in \transrels$ and
$\vars = \set{S^\numcols_1,\ldots,S^\numcols_\numcols}$ for the set of state variables.
\end{definition}

\noindent Both write and read independence are expressed in terms of unordered pairs between
variables and transitions. In a Petri net a transition is both write and read independent from a place when
it neither consumes nor produces a token in that place. If a Petri net is 1-safe, a transition
can always safely produce a token when the transition is enabled and all places always have at most 1 token. Thus in 1-safe Petri nets transitions 
are read independent from places where tokens are produced.

\begin{definition}[Write independence]\label{def:write-independence}
Given a \gls{pts} $\pts = \tuple{\states,\transrels,\inits}$,
transition group $i$ is \emph{write-independent} from state variable $j$, if:
$  \forall \vec{s}, \vec{t} \in \states \colon
    \tuple{s_1, \dotsc, s_j, \dotsc, s_\numcols} \to_i \tuple{t_1, \dotsc, t_j, \dotsc, t_\numrows}
    \implies (s_j = t_j), $
i.e., state variable $j$ is never modified in transition group $i$.
The \emph{write set} is
$\writeset = \set{\set{\to^\numrows_i, S^\numcols_j} \mid \mathop{\to^\numrows_i} \text{ is not write-independent from } S^\numcols_j} \subseteq \binom{\transrels \cup \vars}{2}$.
\end{definition}

\noindent In \Cref{fig:petrinet}, $t_1$ is write independent from $p_1$, but write dependent from $p_4$. 

\begin{definition}[write graph]
Given a \gls{pts} $\pts = \tuple{S, \to, \inits}$ and a write set $\writeset$,
a \emph{write graph} is a partially ordered bipartite graph 
$\writegraph = (\to \cup \mathop{\mathcal{S}}, \writeset, \order{}{})$, where $\order{}{} \subset \to^2 \cup \mathop{\mathcal{S}}^2$.
The biadjacency matrix of the write graph is called the \emph{write matrix}, denoted $\WM \in \set{0,1}^{\sizeof{\to} \times \sizeof{\mathcal{S}}}$,
an element of $\WM$ is $\WME_{ij}$.
\end{definition}

\begin{figure}%
   \vspace{-15pt}
\centering%
\subfloat[Write graph: $\writegraph = \tuple{\to \cup \mathop{\mathcal{S}}, \writeset, \mathcal{A}}$]{{%
\resizebox{.5\textwidth}{!}{%
\begin{tikzpicture}[node distance=2cm,>=stealth',bend angle=45,auto, every node/.style={circle,draw},transform shape]

    \node (T1) {\large{$t_1$}};
    \node[right of=T1] (T2) {\large{$t_2$}};
    \node[right of=T2] (T3) {\large{$t_3$}};
    \node[right of=T3] (T4) {\large{$t_4$}};
    \node[right of=T4] (T5) {\large{$t_5$}};
    \node[right of=T5] (T6) {\large{$t_6$}};
    
    \node[draw=none, below of=T1] (ph) {};
    \node[right of=ph, node distance=1cm] (P1) {\large{$p_1$}};
    \node[right of=P1] (P2) {\large{$p_2$}};
    \node[right of=P2] (P3) {\large{$p_3$}};
    \node[right of=P3] (P4) {\large{$p_4$}};
    \node[right of=P4] (P5) {\large{$p_5$}};
    
    \node[above of=P3, node distance=3cm, draw=none,fill=none] (PT) {\large{Part $\to$}};
    \node[below of=P3, draw=none,fill=none, node distance=1cm] (PP) {\large{Part $\mathcal{S}$}};
    
    \path
    (T1)    edge (P2)
            edge (P4)
            edge (P5)
    (T2)    edge (P2)
            edge (P3)
    (T3)    edge (P2)
            edge (P3)
    (T4)    edge (P1)
            edge (P5)
    (T5)    edge (P1)
            edge (P5)
    (T6)    edge (P1)
            edge (P3)
            edge (P4);
\begin{pgfonlayer}{background}
\draw[style=dashed,thick] \convexpath{PT,T6,T5,T4,T3,T2,T1}{12pt};
\draw[style=dashed,thick] \convexpath{P1,P2,P3,P4,P5,PP}{12pt};
\fill[color=red,opacity=.2] \convexpath{PT,T6,T5,T4,T3,T2,T1}{12pt};
\fill[color=blue,opacity=.2] \convexpath{P1,P2,P3,P4,P5,PP}{12pt};
\end{pgfonlayer}
\end{tikzpicture}
}}}%
\qquad
\subfloat[Write matrix: \WM]{{%
\raisebox{30pt}{%
$\bbordermatrix{%
            & p_1         & p_2       & p_3       & p_4       & p_5       \cr
t_1         & 0    & \mathbf{1}    & 0         & \mathbf{1}         & \mathbf{1}     \cr
t_2         & 0    & \mathbf{1}    & \mathbf{1}         & 0         & 0    \cr 
t_3         & 0    & \mathbf{1}    & \mathbf{1}         & 0         & 0    \cr 
t_4         & \mathbf{1}    & 0    & 0         & 0         & \mathbf{1}       \cr 
t_5         & \mathbf{1}    & 0    & 0         & 0         & \mathbf{1}     \cr 
t_6         & \mathbf{1}    & 0    & \mathbf{1}         & \mathbf{1}         & 0    \cr }
$ }}}%
   \vspace{-5pt}
\caption{Write dependencies}%
\label{fig:deps}%
   \vspace{-15pt}
\end{figure}
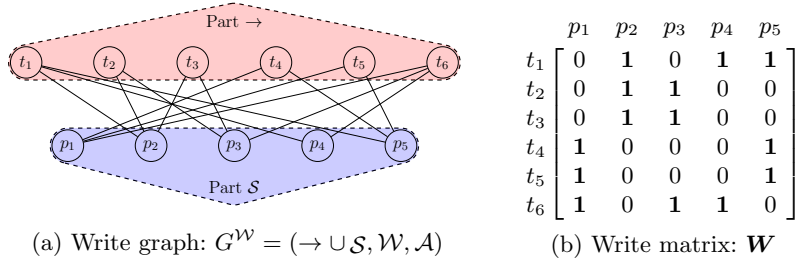

\noindent \Cref{fig:deps} shows the write graph and the write matrix of the Petri net, using the default alphanumeric (partial) order $\mathcal{A}$.

\noindent Analogous to the definitions regarding write dependence we can define read dependence \cite{DBLP:conf/hvc/MeijerKBP14}. 
Furthermore we can define the \emph{combined graph} and \emph{combined matrix} of which the edges and nonzeros
are the elements in the union of the write set and \emph{read set}. Siminiceanu et al. \cite{DBLP:conf/tacas/SiminiceanuC06} 
use a similar concept as the combined matrix to define the WES metric. The concept of the \emph{dependency matrix} in earlier papers
on \ltsmin \cite{DBLP:conf/hvc/MeijerKBP14,blom2008:symbolic} is identical to the combined matrix. In this paper we focus on write
dependence, because only write dependencies can create nodes in decision diagrams. We do not focus
on read dependence, because in Petri nets, there exists no net with only a read dependency on a place (consuming or producing tokens always requires a write dependency).
In languages such as \promela, \dve and \mcrl read dependencies do exist without write dependencies, e.g. as variables on the right hand side of assignments.

\subsection{Saturation and Peak Nodes}\label{ssec:sat}
\Cref{fig:ldd} shows that the final set of reachable states benefits from a good variable ordering.
In general however, the number of nodes in DDs during reachability analysis is much
higher than at the end of reachability. The number of nodes as a function of reachable states
is of parabolic form and its maximum -- and thus limiting factor in terms of memory usage -- is called \emph{peak nodes}.
Although good variable orders allow the number of peak nodes to remain low, an advanced reachability strategy is also mandatory for low peak nodes.
A reachability algorithm can be expressed as a sequence of applying transitions, i.e. let $t_i^=$ be the application
of transition $i$ including the identity. Then
$\textsc{bfs} = \scope{\scope{t_1 \cup \ldots \cup t_{\numrows}}^=}^+$, $\textsc{chaining} = \scope{t_1^= \cdot t_{\ldots}^= \cdot t_{\numrows}^=}^+$ and
$\textsc{sat-like} = ((((t_\numrows^= \cdot t_{\numrows - 1}^=)^+ \cdot t_{\numrows - 2}^=)^+ \cdot t_{\ldots}^=)^+ \cdot t_1^=)^+$.
The \textsc{sat-like} algorithm saturates the bottom of the DD first, to keep peak nodes low, like the \emph{saturation} algorithm by Ciardo et al. \cite{DBLP:journals/sttt/CiardoMS06}.
In LTSmin, the option \texttt{--sat-granularity} allows to group multiple transitions for saturation.
Within a group of transitions, \textsc{chaining} is used. By default, LTSmin creates $\numcols / 10$ transitions groups.

\subsection{Metrics for Undirected Ordered Graphs}\label{ssec:metrics}
The \emph{bandwidth} of a row in a matrix is the maximal distance to the diagonal of a nonzero in that row.
The \emph{span} of a row in a matrix is the distance between the minimal and maximal nonzero.

\begin{definition}[bandwidth]\label{def:bandwidth}
Given a graph $\graph = \tuple{\vertices, \edges, \order{}{}}$,
the \emph{vertex bandwidth}
is a function $\bwdef{\graph} \colon \vertices \to \natnumsetlt{\sizeof{\vertices}}^0$, such that
\begin{tagcases}[v \mapsto]
0 & \text{ if } \neighbors{\graph}{v} = \emptyset, \tag{a}\label{def:bandwidth:1} \\
\max_{w \in \neighbors{\graph}{v}}\abs{\orderfn{\order{}{}}{v} - \orderfn{\order{}{}}{w}} & \text{ otherwise}. \tag{b}\label{def:bandwidth:2}
\end{tagcases}
\end{definition}

\begin{definition}[span]\label{def:span}
Given a graph $\graph = \tuple{\vertices, \edges, \order{}{}}$,
the \emph{vertex span}
is a function $\spndef{\graph} \colon \vertices \to \natnumsetleq{\sizeof{\vertices}}^0$, such that
\begin{tagcases}[v \mapsto]
0 & \text{ if } \neighbors{\graph}{v} = \emptyset, \tag{a}\label{def:span:1} \\
\max_{w \in \neighbors{\graph}{v}}\orderfn{\order{}{}}{w} - \min_{w \in \neighbors{\graph}{v}}\orderfn{\order{}{}}{w} + 1 & \text{ otherwise}. \tag{b}\label{def:span:2} 
\end{tagcases}
\end{definition}

\noindent In \Cref{fig:deps} the bandwidth and span of $t_4$ are $\bw{\writegraph}{t_4} = 3$ and $\spn{\writegraph}{t_4} = 5$.

The \emph{wavefront} of a column in a matrix is the number of rows that have nonzeros within columns smaller or equal than that column. 

\begin{definition}[wavefront]\label{def:wavefront}
Given a graph $\graph = \tuple{\vertices, \edges, \order{}{}}$,
the \emph{vertex frontwidth} or \emph{vertex wavefront} is a function
$\wfdef{\graph} \colon \vertices \to \natnumsetleq{\sizeof{\vertices}}$, such that
$u \mapsto |\set{u} \cup \{v \mid v \in \vertices \land u \neq v \land \exists w \in \vertices \colon \set{w,v} \in \edges \land w \leq u\}|$.
\end{definition}
\noindent In \Cref{fig:deps} the wavefront of $p_3$ is $\wf{\writegraph}{p_3} = 7$, because all transitions depend on a place in $\{p_1,p_2,p_3\}$.
The wavefront of $p_1$ is $\wf{\writegraph}{p_1} = 4$.

\section{Approach}\label{sec:approach}
Our approach consists of transforming the partially ordered adjacency graphs of the dependency matrix
to totally ordered graphs. This is a requirement for the sparse matrix algorithms in \boost and \viennacl.
Then we provide metrics that can measure the quality of orders on a graph level, rather than on
vertex level (\Cref{def:bandwidth,def:span,def:wavefront}). We then explain how nodal ordering algorithms
work on totally ordered graphs and how to apply the resulting permutations to partially ordered graphs.
But first, we prove \Cref{thrm:bandwidth-span} with the triangle inequality. \Cref{thrm:bandwidth-span} shows
that span is limited by twice the bandwidth plus the diagonal. This gives the intuition of why
bandwidth reduction can be used to reduce span. Reducing span is known to be good for variable orderings in symbolic reachability analysis.

\begin{theorem}[bandwidth limits span]\label{thrm:bandwidth-span}
Given a graph $\graph = \tuple{\vertices, \edges, \order{}{}}$, we have
$
\forall v \in \vertices \colon \spn{\graph}{v} \leq 2\cdot\bw{\graph}{v} + 1.
$
\begin{proof}
Take an arbitrary $v \in \vertices$, we have two cases for $v$:
\begin{mycase}[$\neighbors{}{v} = \emptyset$]
By \Cref{def:bandwidth}\ref{def:bandwidth:1} and \ref{def:span}\ref{def:span:1}
we have $0 \leq 2 \cdot 0 + 1\iff \spn{\graph}{v} \leq 2 \cdot \bw{\graph}{v} + 1$.
\end{mycase}
\begin{mycase}[$\sizeof{\neighbors{}{v}} > 0$]
We have two symmetric cases.
\begin{mysubcase}[$\abs{\min_{w \in \neighbors{}{v}} \orderfn{}{w} - \orderfn{}{v}} \geq \abs{\orderfn{}{v} - \max_{w \in \neighbors{}{v}} \orderfn{}{w}}$]\label{thrm:bandwidth-span:sub}
\begin{align}
    &\spn{\graph}{v} = \max_{w \in \neighbors{}{v}} \orderfn{}{w} - \min_{w \in \neighbors{}{v}} \orderfn{}{w} + 1
    = \abs{\max_{w \in \neighbors{}{v}} \orderfn{}{w} - \min_{w \in \neighbors{}{v}} \orderfn{}{w}} + 1 \notag \\
    &= \abs{\max_{w \in \neighbors{}{v}} \orderfn{}{w} - \orderfn{}{v} + \orderfn{}{v} - \min_{w \in \neighbors{}{v}} \orderfn{}{w}} + 1 \notag \\
    &\leq \abs{\max_{w \in \neighbors{}{v}} \orderfn{}{w} - \orderfn{}{v}} + \abs{\orderfn{}{v} - \min_{w \in \neighbors{}{v}} \orderfn{}{w}} + 1
    \leq 2 \cdot \abs{\orderfn{}{v} - \min_{w \in \neighbors{}{v}} \orderfn{}{w}} + 1 \notag \\
    &= 2 \cdot \max_{w \in \neighbors{}{v}} \abs{\orderfn{}{v} - \orderfn{}{w}} + 1 = 2 \cdot \bw{\graph}{v} + 1. \notag
\end{align}
\end{mysubcase}
\begin{mysubcase}[$\abs{\orderfn{}{v} - \max_{w \in \neighbors{}{v}} \orderfn{}{w}} \geq \abs{\min_{w \in \neighbors{}{v}} \orderfn{}{w} - \orderfn{}{v}}$]~\\
Symmetric to \Cref{thrm:bandwidth-span:sub}.
\end{mysubcase}
\end{mycase}
\noindent Since $v$ is arbitrary we are done.
\end{proof}
\end{theorem}

\subsection{Total Orders for Dependency Graphs}

\begin{wrapfigure}{r}{0.41\textwidth}
\vspace{-30pt}
\centering
\resizebox{.42\textwidth}{!}{
$\bbordermatrix{
    & t_1 & t_2 & t_3 & t_4 & t_5 & t_6 & p_1 & p_2 & p_3 & p_4 & p_5 \cr
t_1 & 0   & 0   & 0   & 0   & 0   & 0   & 0   & \mathbf{1}   & 0   & \mathbf{1}   & \mathbf{1}   \cr
t_2 & 0   & 0   & 0   & 0   & 0   & 0   & 0   & \mathbf{1}   & \mathbf{1}   & 0   & 0   \cr 
t_3 & 0   & 0   & 0   & 0   & 0   & 0   & 0   & \mathbf{1}   & \mathbf{1}   & 0   & 0   \cr 
t_4 & 0   & 0   & 0   & 0   & 0   & 0   & \mathbf{1}   & 0   & 0   & 0   & \mathbf{1}   \cr 
t_5 & 0   & 0   & 0   & 0   & 0   & 0   & \mathbf{1}   & 0   & 0   & 0   & \mathbf{1}   \cr 
t_6 & 0   & 0   & 0   & 0   & 0   & 0   & \mathbf{1}   & 0   & \mathbf{1}   & \mathbf{1}   & 0   \cr
p_1 & 0   & 0   & 0   & \mathbf{1}   & \mathbf{1}   & \mathbf{1}   & 0   & 0   & 0   & 0   & 0   \cr
p_2 & \mathbf{1}   & \mathbf{1}   & \mathbf{1}   & 0   & 0   & 0   & 0   & 0   & 0   & 0   & 0   \cr
p_3 & 0   & \mathbf{1}   & \mathbf{1}   & 0   & 0   & \mathbf{1}   & 0   & 0   & 0   & 0   & 0   \cr
p_4 & \mathbf{1}   & 0   & 0   & 0   & 0   & \mathbf{1}   & 0   & 0   & 0   & 0   & 0   \cr
p_5 & \mathbf{1}   & 0   & 0   & \mathbf{1}   & \mathbf{1}   & 0   & 0   & 0   & 0   & 0   & 0   \cr}
$
}
\vspace{-5pt}
\caption{Symmetrized write matrix}
\label{fig:write-matrix-sym}
\vspace{-20pt}
\end{wrapfigure}

The bandwidth and wavefront reduction algorithms discussed in this paper can only be run on the
underlying graphs of symmetric matrices. The method of symmetrizing suggested by Reid et al. \cite{DBLP:journals/siammax/ReidS06} is 
to create the matrix \amtrx from \autoref{def:matrix}, of which an example is shown in \autoref{fig:write-matrix-sym}.
The key difference between the symmetric matrix in \autoref{fig:write-matrix-sym} and asymmetric matrix in \autoref{fig:deps} 
is that symmetric matrix implies a total order on the underlying graph, while the asymmetric matrix implies a partial order.
The total order implied is indeed $T = t_1 < t_2 < t_3 < t_4 < t_5 < t_6 < p_1 < p_2 < p_3 < p_4 < p_5$, while the partial order is
$P = t_1 < t_2 < t_3 < t_4 < t_5 < t_6 \cup p_1 < p_2 < p_3 < p_4 < p_5$. Technically we have two undirected ordered graphs of the form
$G = \tuple{\vertices, \edges, T}$ and $H = \tuple{\vertices, \edges, P}$. On both graphs we can define aggregation and normalization
functions, which allow us to measure the quality on a graph level, instead of vertex level. A key observation
on this approach is that we can actually choose many total orders as long as the partial order remains the same (i.e. $P \subseteq T$). 
This implies different results for computing bandwidth, span and wavefront on $G$, but not on $H$.
Although \Cref{def:bandwidth,def:wavefront} allows measuring bandwidth and wavefront on $H$, it is currently not known
whether these metrics are actually appropriate for symbolic model checking.
We therefore compute those two metrics on $G$, because this is known to appropriate for sparse matrix solvers and is already supported
by \boost and \viennacl.
To see whether bandwidth and wavefront computed on $H$ is appropriate for symbolic reachability analysis requires the same approach taken by Ciardo et al. in \cite{DBLP:conf/tacas/SiminiceanuC06},
where many samples of a metric are generated of a specific model and analyzed.
The results from our benchmark do show however, that metrics computed on $G$ and $H$ are related.

\subsection{Aggregation and Normalization for graph metrics}

\begin{figure}
\vspace{-20pt}
\begin{tabular}{| l | c | c | c | c |}
\hline
\textbf{name} & \textbf{aggregation} & \textbf{normalization} & \textbf{value} \\ \hline
Bandwidth & $\max_{v \in \vertices}\bw{G}{v}$ & $\numrows + \numcols$ & $10~(.91)$ \\ \hline
Profile & $\sum_{v \in \vertices}\bw{G}{v}$ & $\scope{\numrows + \numcols}^2$ & $87~(.72)$ \\ \hline
Span & $\sum_{v \in \vertices}\spn{G}{v}$ & $\scope{\numrows + \numcols}^2$ & $44~(.36)$ \\ \hline
Average wavefront & $\sum_{v \in \vertices}\wf{G}{v} / (\numrows + \numcols)$ & $\numrows + \numcols$ & $4.3~(.39)$ \\ \hline \hline
Event Span & $\displaystyle \sum_{\to \in \transrels}\spn{H}{\to}$ & $\numrows \cdot \numcols$ & $22~(.73)$ \\ \hline
Weighted Event Span & $\displaystyle \sum_{\to \in \transrels} \spn{H}{\to} \cdot \frac{\numcols - \min_{v \in \neighbors{}{\to}}\orderfn{P}{v}}{\numcols / 2}$ & $\numrows \cdot \numcols$ & $41~(1.4)$ \\ \hline
\end{tabular}
\vspace{-5pt}
\caption{Aggregation and normalization}
\label{fig:aggr-norm}
\vspace{-20pt}
\end{figure}

\noindent\Cref{fig:aggr-norm} shows a list of aggregated and normalized metrics for bandwidth, span and wavefront we compute in our benchmarks.
The first four metrics are from the literature on sparse matrix solvers. The last two are by Siminiceanu et al. \cite{DBLP:conf/tacas/SiminiceanuC06}.
Given a \gls{pts} $\pts = \tuple{\states, \transrels, \inits}$, all three type of metrics can be computed on the totally ordered write graph 
\graph in \boost and \viennacl 
or the partially ordered write graph 
$H$ in \ltsmin.
The column \emph{normalization} indicates the factor which should be divided by to obtain a number between zero and one, except for WES.
The column \emph{value} contains the computed values (and normalized values) for the Petri net with the total order from
\Cref{fig:write-matrix-sym} (graph $G$) and the partial order from \Cref{fig:deps} (graph $H$).
Note that the aggregation functions for \tg are the same as for \graph, but are not computed here.

\subsection{Nodal Ordering}\label{ssec:nodal-ordering}
Cuthill Mckee \cite{Cuthill:1969:RBS:800195.805928} is a nodal ordering algorithm for bandwidth reduction. 
The algorithm is a simple breadth-first graph traversal algorithm that visits neighbors of a vertex in increasing order of degree.
For picking a starting vertex, there are several options. The implementation in the Boost graph library
uses a pseudo-peripheral pair heuristic \cite{George:1979:IPN:355841.355845} for each disconnected component.
A simpeler approach is to take the smallest vertex with minimum degree. Using the latter option, the order in which Cuthill McKee
visits vertices is $\order{}{\permdef} = t_2 < p_2 < p_3 < t_3 < t_1 < t_6 < p_4 < p_5 < p_1 < t_4 < t_5$.
The reordered write graph, (symmetrized) write matrix and metrics of the Petri net are shown in \Cref{fig:deps:reord}.
The \emph{band} and low \emph{event span} in \Cref{fig:deps:reord:write-matrix} is clearly visible. Note that indeed
the partial order for the reordered write graph is $\order{\vars}{\permdef} \cup \order{\transrels}{\permdef}$.
If a nodal ordering algorithm is run on the total graph of the write graph to obtain $O^{T\permdef}$, the partial order for the write graph is simply
$\order{\vars}{T\permdef} \cup \order{\transrels}{T\permdef}$.

\begin{figure}[t]%
\vspace{-20pt}
\centering%
\subfloat[Write graph: $\tuple{\transrels \cup \vars, \writeset, \order{\vars}{\permdef} \cup \order{\transrels}{\permdef}}$]{{%
\resizebox{.5\textwidth}{!}{%
\centering
\begin{tikzpicture}[node distance=2cm,>=stealth',bend angle=45,auto, every node/.style={circle,draw}]

    \node (T2) {\large{$t_2$}};
    \node[right of=T2] (T3) {\large{$t_3$}};
    \node[right of=T3] (T1) {\large{$t_1$}};
    \node[right of=T1] (T6) {\large{$t_6$}};
    \node[right of=T6] (T4) {\large{$t_4$}};
    \node[right of=T4] (T5) {\large{$t_5$}};
    
    \node[draw=none, below of=T2] (ph) {};
    \node[right of=ph, node distance=1cm] (P2) {\large{$p_2$}};
    \node[right of=P2] (P3) {\large{$p_3$}};
    \node[right of=P3] (P4) {\large{$p_4$}};
    \node[right of=P4] (P5) {\large{$p_5$}};
    \node[right of=P5] (P1) {\large{$p_1$}};
    
    \node[above of=P4, node distance=3cm, draw=none,fill=none] (PT) {\large{Part $\transrels$}};
    \node[below of=P4, draw=none,fill=none, node distance=1cm] (PP) {\large{Part $\vars$}};
    
    \path
    (T1)    edge (P2)
            edge (P4)
            edge (P5)
    (T2)    edge (P2)
            edge (P3)
    (T3)    edge (P2)
            edge (P3)
    (T4)    edge (P1)
            edge (P5)
    (T5)    edge (P1)
            edge (P5)
    (T6)    edge (P1)
            edge (P3)
            edge (P4);
            
\begin{pgfonlayer}{background}
\draw[style=dashed,thick] \convexpath{PT,T5,T4,T6,T1,T3,T2}{12pt};
\draw[style=dashed,thick] \convexpath{PP,P2,P3,P4,P5,P1}{12pt};
\fill[color=red,opacity=.2] \convexpath{PT,T5,T4,T6,T1,T3,T2}{12pt};
\fill[color=blue,opacity=.2] \convexpath{PP,P2,P3,P4,P5,P1}{12pt};
\end{pgfonlayer}
           
\end{tikzpicture}
} }}%
\qquad
\subfloat[Write matrix]{\label{fig:deps:reord:write-matrix}{%
\raisebox{35pt}{%
\resizebox{.3\textwidth}{!}{%
$\bbordermatrix{
            & p_2  & p_3  & p_4  & p_5  & p_1  \cr
t_2         & \mathbf{1}    & \mathbf{1}    & 0    & 0    & 0    \cr
t_3         & \mathbf{1}    & \mathbf{1}    & 0    & 0    & 0    \cr 
t_1         & \mathbf{1}    & 0    & \mathbf{1}    & \mathbf{1}    & 0    \cr 
t_6         & 0    & \mathbf{1}    & \mathbf{1}    & 0    & \mathbf{1}    \cr 
t_4         & 0    & 0    & 0    & \mathbf{1}    & \mathbf{1}    \cr 
t_5         & 0    & 0    & 0    & \mathbf{1}    & \mathbf{1}    \cr }
$ }}}}%
\vspace{-10pt}
\subfloat[Metrics]{{%
\raisebox{7pt}{%
\resizebox{.3\textwidth}{!}{%
\begin{tabular}{|l|c|}
\hline
\textbf{Bandwidth} & $3~(.27)$ \\ \hline
\textbf{Profile} & $40~(.33)$ \\ \hline
\textbf{Span} & $48~(.40)$ \\ \hline
\textbf{Avg wavefront} & $3.2~(.29)$ \\ \hline
\textbf{ES} & $16~(.53)$ \\ \hline
\textbf{WES} & $26~(.87)$ \\ \hline
\end{tabular}}}}}%
\qquad
\subfloat[Symmetrized write matrix]{{%
\raisebox{9pt}{%
\resizebox{.4\textwidth}{!}{
$\bbordermatrix{
    & t_2 & p_2 & p_3 & t_3 & t_1 & t_6 & p_4 & p_5 & p_1 & t_4 & t_5 \cr
t_2 & 0   & \mathbf{1}   & \mathbf{1}   & 0   & 0   & 0   & 0   & 0   & 0   & 0   & 0   \cr
p_2 & \mathbf{1}   & 0   & 0   & \mathbf{1}   & \mathbf{1}   & 0   & 0   & 0   & 0   & 0   & 0   \cr 
p_3 & \mathbf{1}   & 0   & 0   & \mathbf{1}   & 0   & \mathbf{1}   & 0   & 0   & 0   & 0   & 0   \cr 
t_3 & 0   & \mathbf{1}   & \mathbf{1}   & 0   & 0   & 0   & 0   & 0   & 0   & 0   & 0   \cr 
t_1 & 0   & \mathbf{1}   & 0   & 0   & 0   & 0   & \mathbf{1}   & \mathbf{1}   & 0   & 0   & 0   \cr 
t_6 & 0   & 0   & \mathbf{1}   & 0   & 0   & 0   & \mathbf{1}   & 0   & \mathbf{1}   & 0   & 0   \cr
p_4 & 0   & 0   & 0   & 0   & \mathbf{1}   & \mathbf{1}   & 0   & 0   & 0   & 0   & 0   \cr
p_5 & 0   & 0   & 0   & 0   & \mathbf{1}   & 0   & 0   & 0   & 0   & \mathbf{1}   & \mathbf{1}   \cr
p_1 & 0   & 0   & 0   & 0   & 0   & \mathbf{1}   & 0   & 0   & 0   & \mathbf{1}   & \mathbf{1}   \cr
t_4 & 0   & 0   & 0   & 0   & 0   & 0   & 0   & \mathbf{1}   & \mathbf{1}   & 0   & 0   \cr
t_5 & 0   & 0   & 0   & 0   & 0   & 0   & 0   & \mathbf{1}   & \mathbf{1}   & 0   & 0   \cr}
$} }}}%
\vspace{-5pt}
\caption{Reordered Petri net}%
\label{fig:deps:reord}%
\end{figure}


\begin{figure}[t]
\vspace{-5pt}
\begin{tabular}{|l|c|l|c|c|c|}
\hline
\textbf{Algorithm} & \textbf{Package} & \textbf{Time complexity} & \textbf{Type} & \textbf{Graph} \\ \hline
Cuthill McKee & \multirow{3}{*}{\boost} & $\bigo{\maxdeg \cdot \log \maxdeg \cdot \sizeof{\vertices}}$ & bandwidth & \multirow{3}{*}{\parbox{2.1cm}{totally ordered undirected graph}} \\ \cline{1-1} \cline{3-4}
King \cite{NME:NME1620020406} & & $\bigo{\maxdeg^2 \cdot \log \maxdeg \cdot \sizeof{\edges}}$ & bandwidth, profile & \\ \cline{1-1} \cline{3-4} 
Sloan & & $\bigo{\maxdeg \cdot \log \maxdeg \cdot \sizeof{\vertices}}$ & profile, wavefront & \\ \hline 
Cuthill McKee & \multirow{3}{*}{\viennacl} & n/a & bandwidth & \multirow{3}{*}{\parbox{2.1cm}{totally ordered directed graph}} \\ \cline{1-1} \cline{3-4}
adv. Cuthill McKee & & n/a & bandwidth & \\ \cline{1-1} \cline{3-4}
GPS \cite{gibbs1976algorithm} & & n/a & bandwidth, profile & \\ \hline
Column Swap & \ltsmin & $\bigo{\numrows^2 \cdot \numcols^4}$ & event span & \multirow{2}{*}{\parbox{2.1cm}{asymmetric matrix}} \\  \cline{1-4}
\multicolumn{4}{|l|}{\pbox{\textwidth}{\textbf{Notation}: $\maxdeg = \max_{v \in V}\sizeof{\neighbors{}{v}}$ (maximum degree)}} & \\ \hline
\end{tabular}
\vspace{-5pt}
\caption{List of reordering algorithms}
\label{fig:list:alg}
\vspace{-15pt}
\end{figure}

\Cref{fig:list:alg} lists all the reordering algorithms that we have considered and their attributes.
There are four categories of algorithms, those that reduce bandwidth, bandwidth and profile,
reduce wavefront and profile, and those that reduce event span. In both \boost and \viennacl the Cuthill McKee algorithm is implemented,
which use an undirected and directed graph respectively, as datastructures. \Cref{sec:results} confirms that the Cuthill McKee implementation
differ in both tools. The GPS algorithm is only implemented in \viennacl and the time complexity
of algorithms in \viennacl is not precisely known, but must be in the order of similar BFS algorithms.
One special algorithm in the list is the column swap algorithm, which is a heuristic algorithm in \ltsmin.
Its key feature is that column swap generates permutations that give low event span. In general it produces good ES, but is unpractical for large matrices.

\begin{figure}
\vspace{-35pt}
\centering
\subfloat[Philosophers-20.pnml]{\label{fig:philo-20}{%
\resizebox{.5\linewidth}{!}{%
\begin{tabular}{c c}
\Large{None} & \Large{Cuthill McKee} \\
\fbox{\includegraphics[width=.4\textwidth]{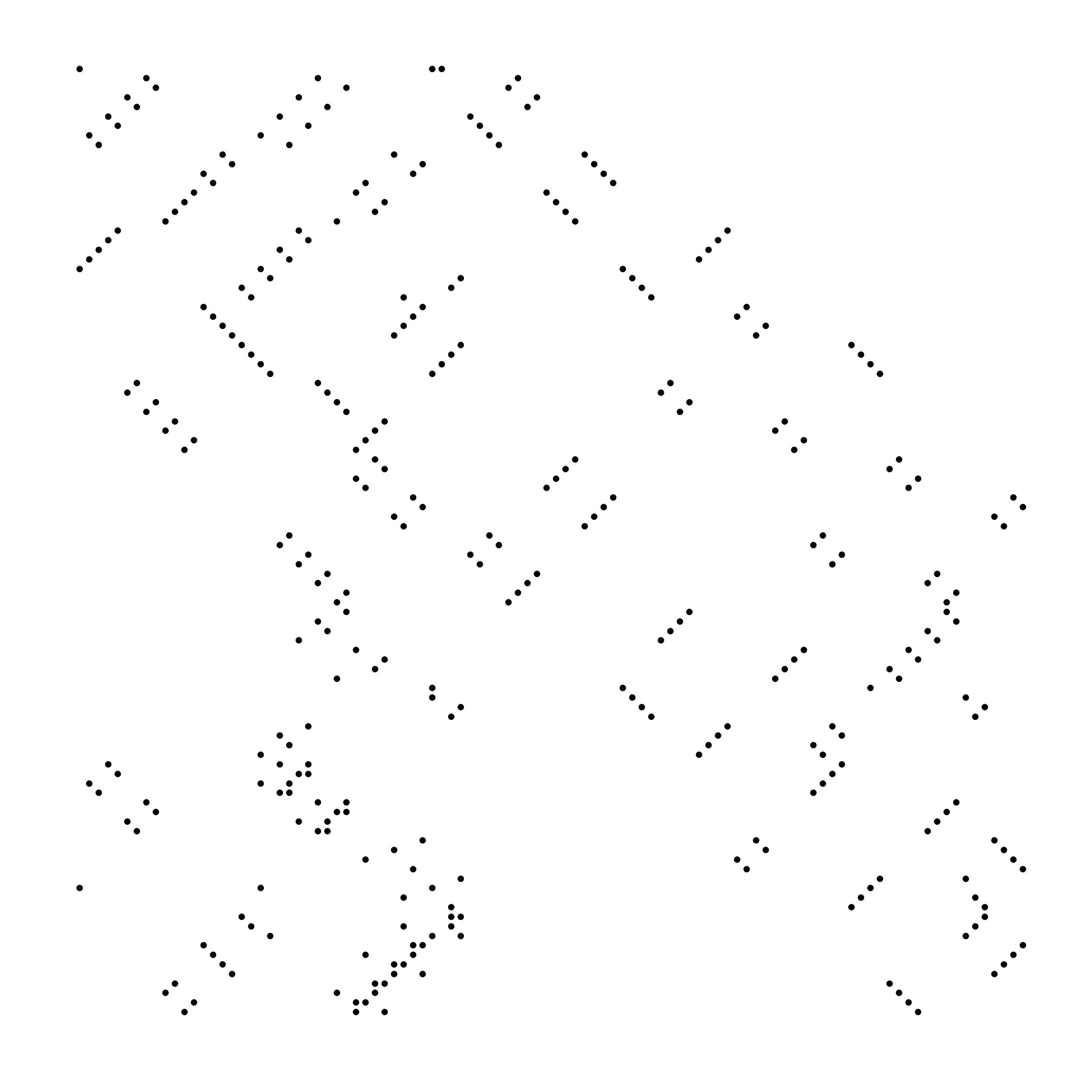}} & \fbox{\includegraphics[width=.4\textwidth]{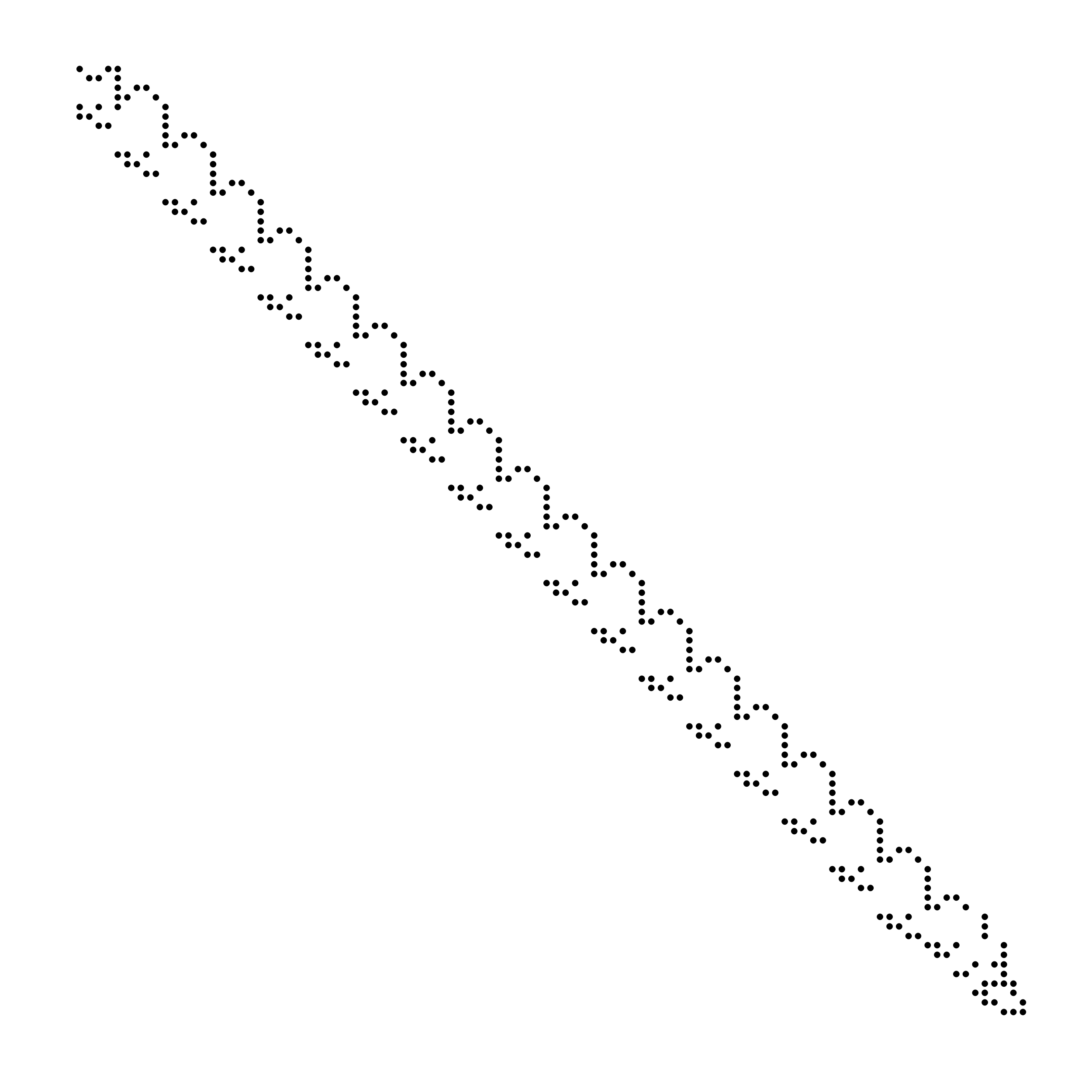}} \\
\multicolumn{1}{l}{\large{~WES:\qquad~~1.0}} & \large{0.08} \\
\end{tabular}
}}}%
\subfloat[Vasy2003.pnml]{\label{fig:Vasy2003}{%
\resizebox{.5\linewidth}{!}{%
\raisebox{3pt}{%
\begin{tabular}{c c c}
\Large{None} & \Large{Sloan} & \Large{GPS} \\
    \fbox{\includegraphics[width=.29\textwidth]{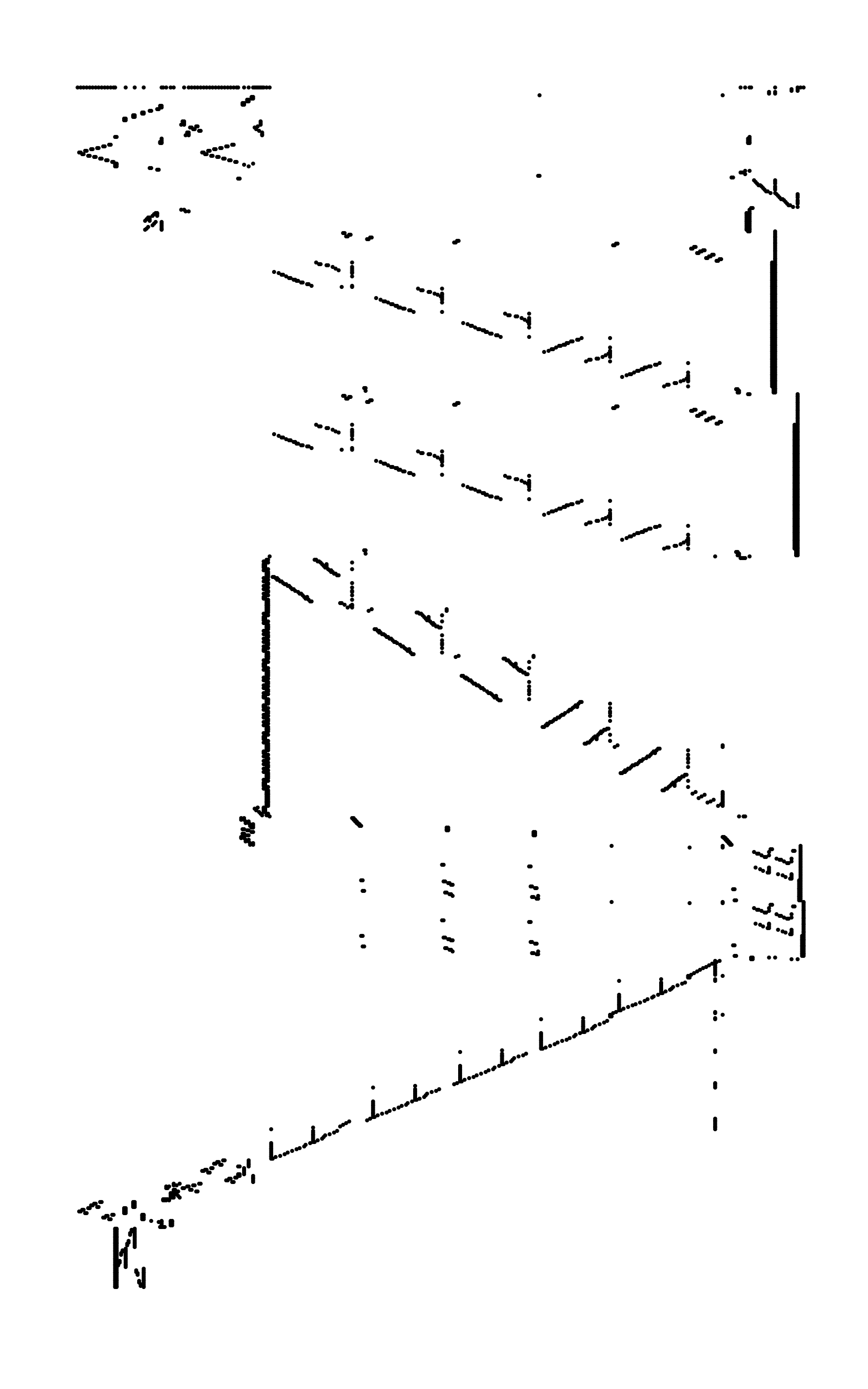}} &
    \fbox{\includegraphics[width=.29\textwidth]{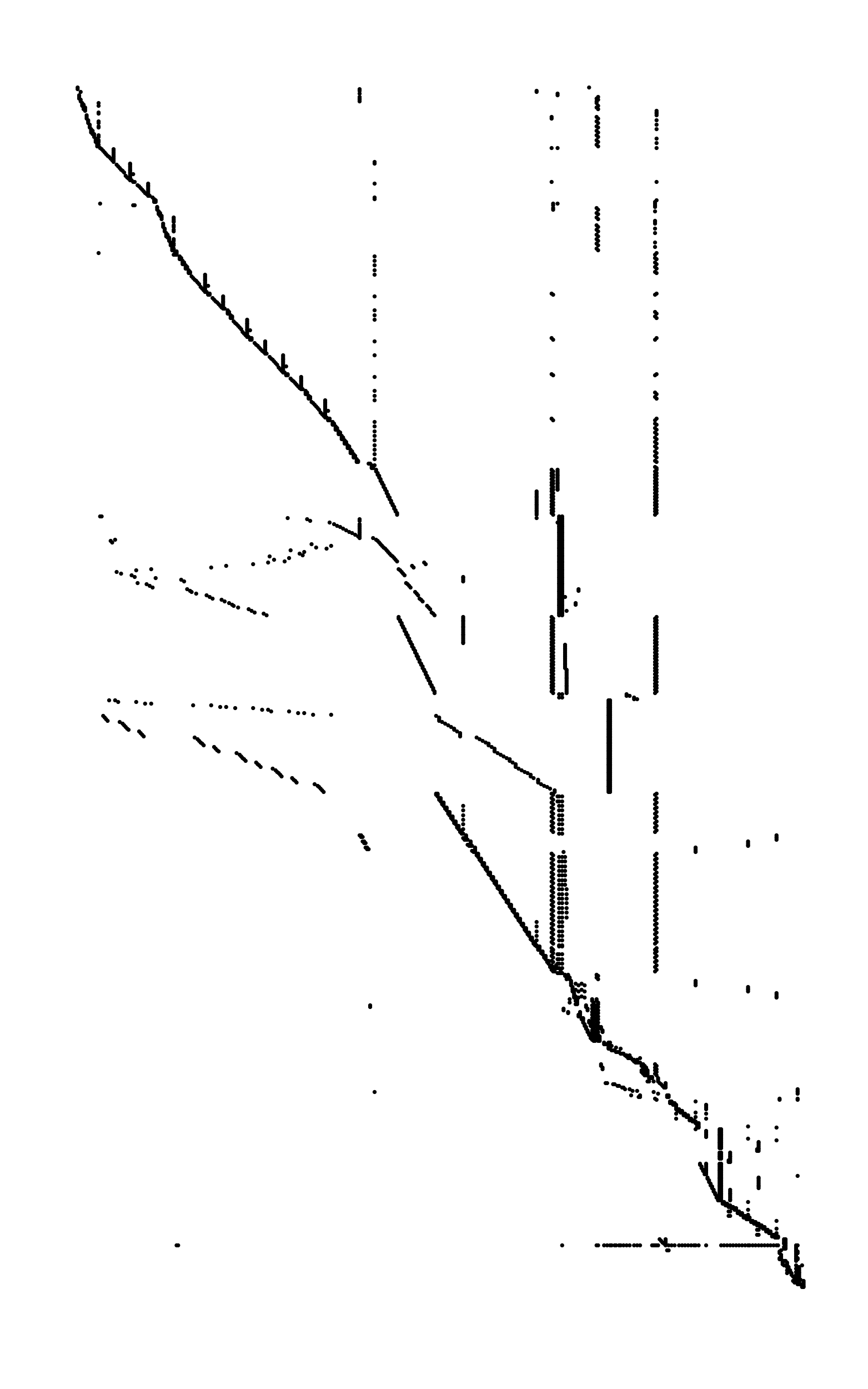}} &
    \fbox{\includegraphics[width=.29\textwidth]{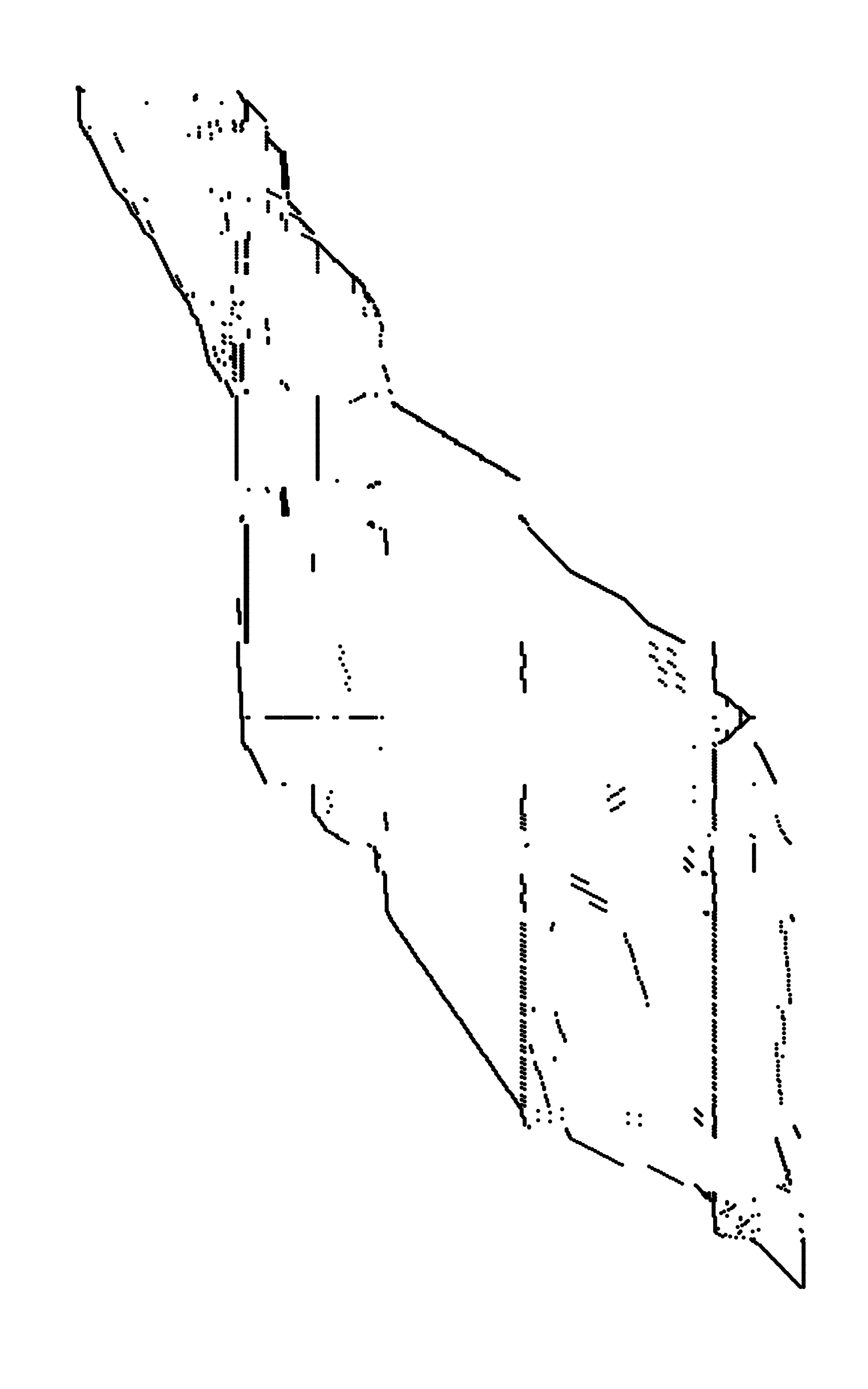}} \\
\large{0.311} & \large{0.375} & \large{0.474} \\
\end{tabular}
}}}}%
\vspace{-5pt}
\caption{Example reorderings}
\vspace{-20pt}
\end{figure}

\Cref{fig:philo-20,fig:Vasy2003} show what dependency matrices can look like when reordered with
some algorithms from \autoref{fig:list:alg} including their weighted event span. The first two matrices are of a model with 20 dining philosophers,
one of the best results achieved in our benchmarks. Even on instances with 5000 philosophers (\emph{25.000 variables}) we get very small event span.
The order is computed within milliseconds and
the resulting number of peak nodes is several orders of magnitude smaller than without reordering. The matrices from the Vasy2003 model
show the more typical structure of dependency matrices, e.g. the \emph{band} the GPS algorithm produces is clearly visible. The Sloan
algorithm produces the best order for Vasy2003, probably because of the large concentration of many nonzeros on the bottom end of the diagonal.
We believe this concentration of nonzeros is produced due to wavefront reduction and is beneficial to the saturation algorithm.

When benchmarking with \ltsmin, we actually need the banded structure of permuted matrices to appear from the top right to bottom left.
This can be done by inverting the permutation of transitions (rows) or permutation of variables (columns), these operations are known as
\emph{horizontal flip} and \emph{vertical flip} respectively. A flip operation is necessary, because it influences the result of the
\textsc{chaining} part in \textsc{sat-like}.
It is yet unclear which flip operation works best, thus in the benchmark we try both.

\section{Results}\label{sec:results}
We have benchmarked the sparse matrix ordering algorithms on 361 different Petri net models
from the 2015 model checking contest\footnote{\url{http://mcc.lip6.fr}}. Reproduction instructions can be found online\footnote{\url{https://github.com/utwente-fmt/BW-TACAS-2016}}.
The benchmark consists of 6 different algorithms from \Cref{fig:list:alg}, computed on the write graph and its total graph.
Additionally, the column swap algorithm is run, as well as no reordering algorithm.
Furthermore we added the options \emph{Horizontal Flip} and \emph{Vertical Flip} which invert the permutation on rows and columns, respectively. 
We do not perform a flip operation when no reordering is done.
Our benchmark consists of 53 categories, but we show only 27 $(= ((6 \cdot 2) + 1) \cdot 2 + 1)$ categories, since we omit the results of the combined graph.
In total we did 114798 $(= 53 \cdot 361 \cdot (1 + 5))$ experiments, which contains 1 run for each category to obtain statistics 
(e.g. peak nodes and metrics) and 5 runs to measure time. Our results include
only those models, which all categories were able to compute within 30 minutes and 4 GB of memory. This resulted in 110 usable models.
We ran 1 experiment on 1 core on 12 single and 32 dual
socket AMD Opteron 4386 processors, with 64 GB of memory and Ubuntu 14.04 LTS. To be able to complete the entire
benchmark in two days we ran 608 experiments simultaneously. The advantage of this approach is that
we were able to gather lots of data on different algorithms and models. The disadvantage is that
time measurements are less reliable. However, the measurement of peak nodes is more important than time measurements.

\begin{figure}
\vspace{-10pt}
    \includegraphics[width=\textwidth]{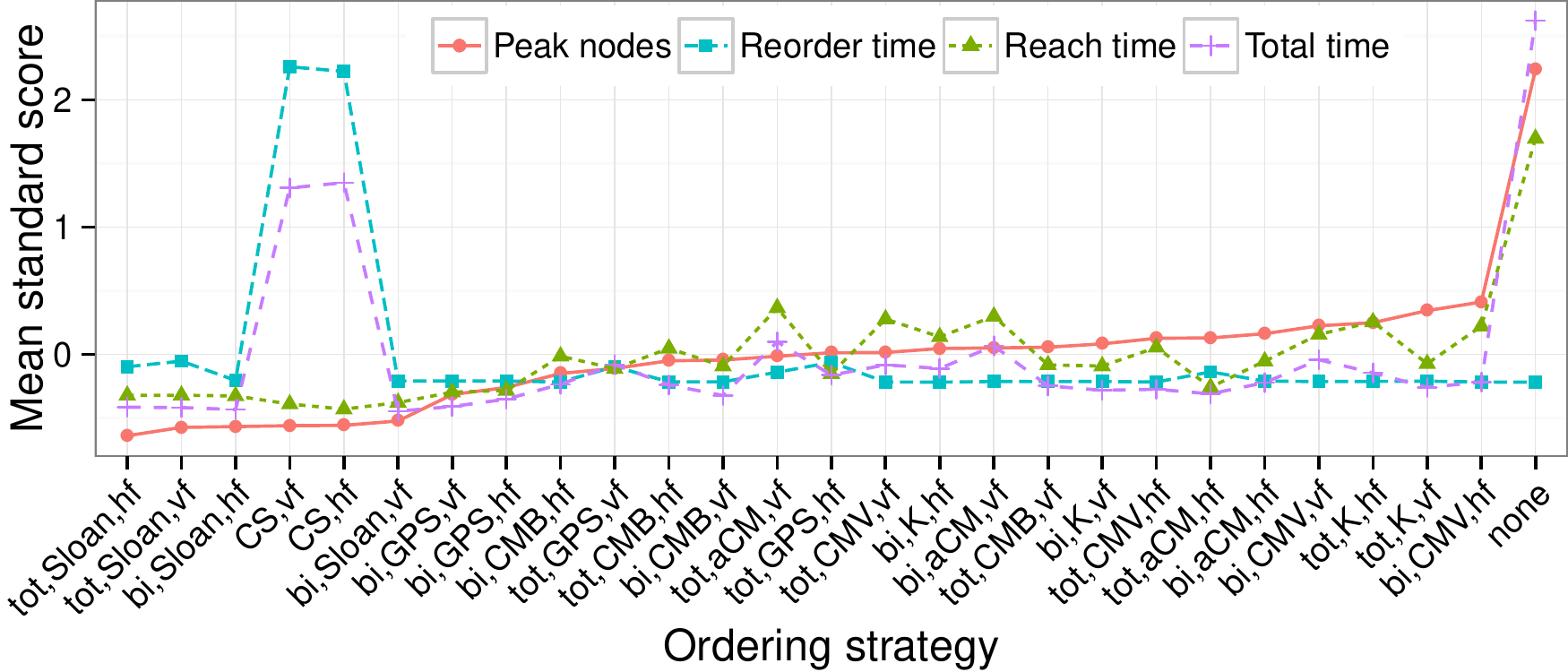}
\vspace{-15pt}
    \caption{Performance of reordering algorithms}
    \label{fig:performance}
\vspace{-20pt}
\end{figure}

\noindent\Cref{fig:performance} shows which algorithm produces the best result. The results are ordered
in ascending number of peak nodes. The category \emph{tot,Sloan,hf} means we performed the \textbf{Sloan}
algorithm on the \textbf{tot}al graph of the write graph. Then we computed all metrics in \Cref{fig:aggr-norm}, performed
a \textbf{h}orizontal \textbf{f}lip and ran the reachability algorithm. The \emph{Mean Standard Score} (MSS), also known as mean z-score, indicates
how well a category performs relative to the mean ($\mu$) observed value of models divided by the standard deviation ($\sigma$).
For example, the \emph{tot,Sloan,hf} category has the lowest score for peak nodes (it appears on the left),
while \emph{none} has the lowest score for reorder time (since no reordering is done). More precisely, 
let $C$ be the set of categories, $m$ a metric such as peak nodes of a category and a model,
and $P$ the set of models with completed runs, the MSS of a category $c \in C$ is: 
$\sum_{p \in P}\frac{m(p,c) - \mu_{c' \in C} m(p,c')}{\sigma_{c' \in C} m(p,c')} / \sizeof{P}$.
Other abbreviations in \autoref{fig:performance} that require explanation are: 
CS = \textbf{C}olumn \textbf{S}wap, bi = \textbf{bi}partite graph, CMB = \textbf{C}uthill \textbf{M}cKee in \textbf{B}oost,
aCM = \textbf{a}dvanced \textbf{C}uthill \textbf{M}cKee, K = \textbf{K}ing and CMV = \textbf{C}uthill \textbf{M}cKee in \textbf{V}iennaCL.

\begin{figure}
\vspace{-20pt}
    \includegraphics[width=\textwidth]{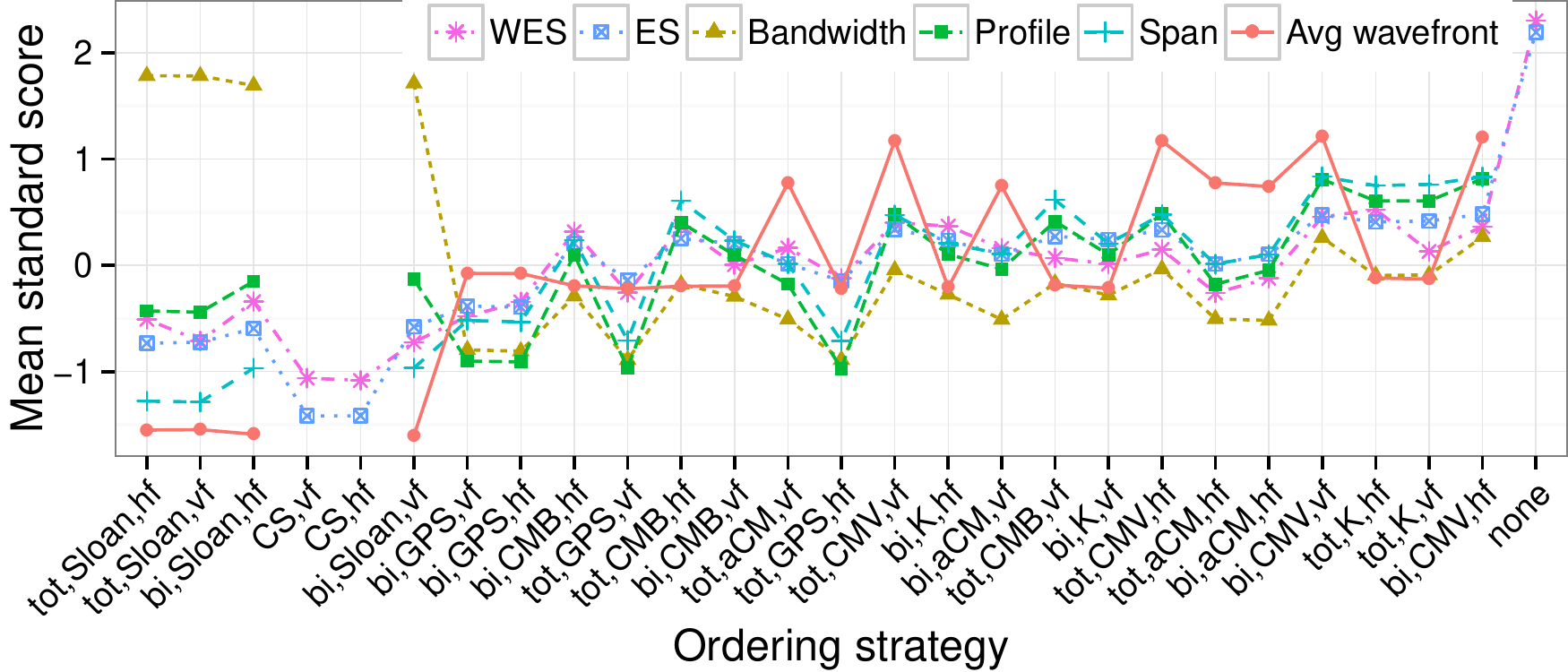}
\vspace{-15pt}
    \caption{Metrics produced with reordering algorithms}
    \label{fig:metrics}
\vspace{-20pt}
\end{figure}

\noindent\Cref{fig:metrics} shows the MSS for the matrix metrics with the same order for categories as \autoref{fig:performance}.
The totally ordered graph is not created with \emph{column swap} and \emph{none}, thus bandwidth, profile, span and wavefront are omitted for those categories.
Furthermore the $\mu$ and $\sigma$ for bandwidth, profile, span and wavefront are computed per graph type, i.e. for all bipartite
graphs with total order and for all total graphs, because those two graph types have different number of vertices and edges.
A few observations can be made, such as that the WES metric is a good predictor for the number of peak nodes in our benchmark; the WES increases with the
number of peak nodes.
It is a good idea to perform any reordering algorithm rather than none. The \emph{Sloan} algorithm performs well for all metrics, but bandwidth.
The GPS algorithm performs second best to Sloan. The wavefront is higher for GPS algorithms and can thus explain why Sloan performs better than GPS in terms of peak nodes.

\begin{figure}
\vspace{-20pt}
\subfloat[Reachability time (seconds)]{
    \includegraphics[width=.48\textwidth]{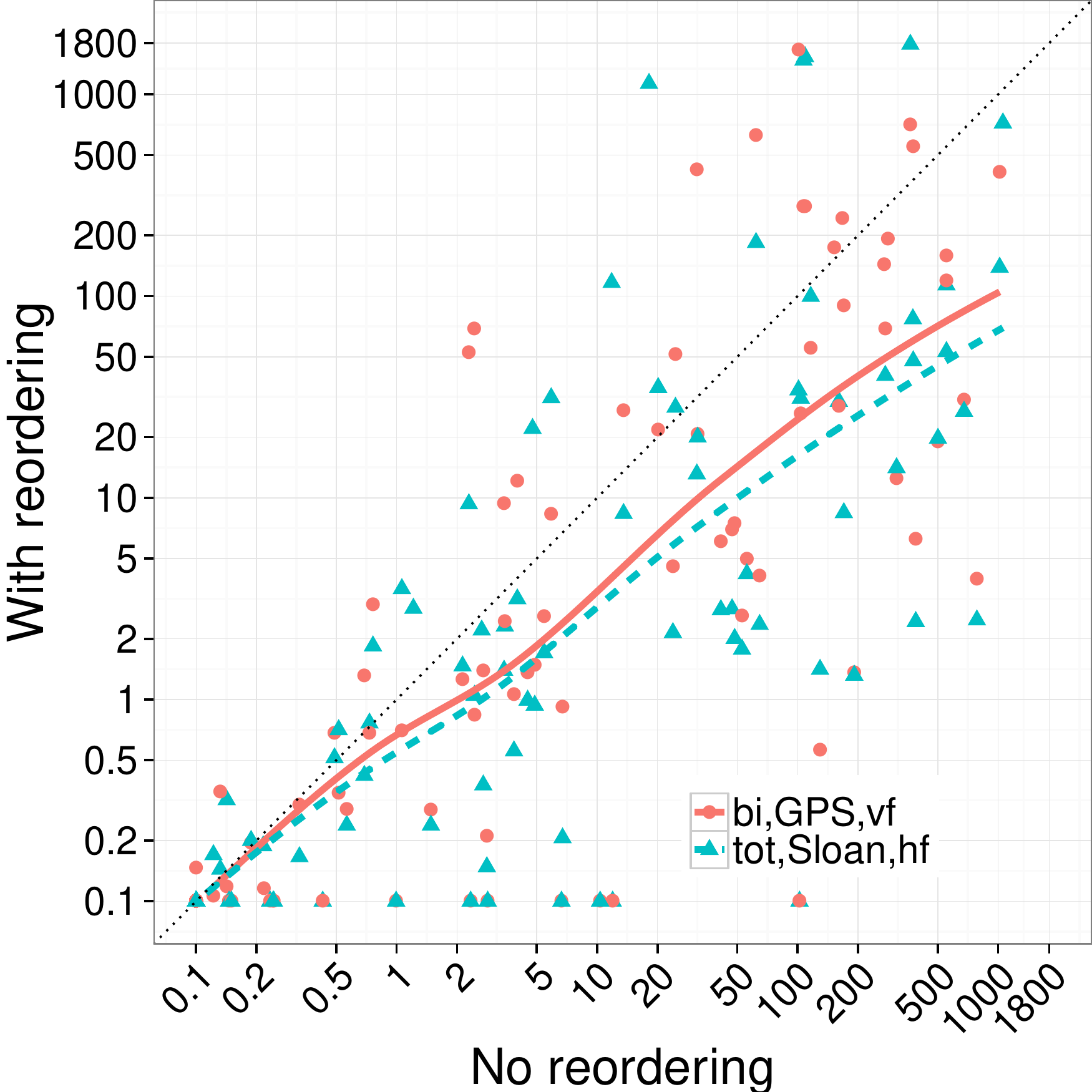}
    \label{fig:reach-time}
}%
\subfloat[Peak nodes]{
    \includegraphics[width=.48\textwidth]{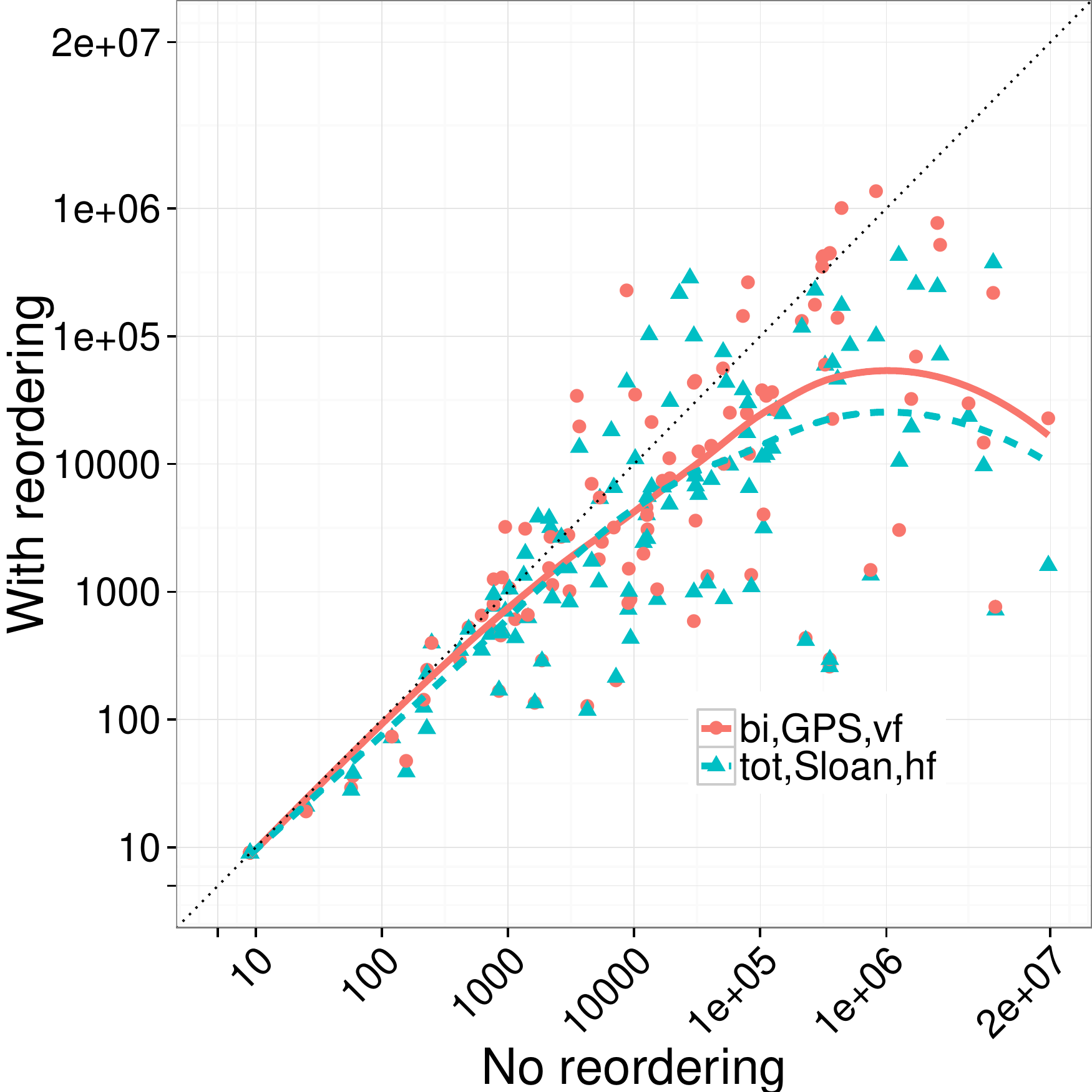}
    \label{fig:peak-nodes}
}
\vspace{-5pt}
\caption{Comparison of Sloan and GPS to no reordering}
\label{fig:comparison}
\vspace{-20pt}
\end{figure}

\Cref{fig:comparison} shows how well the Sloan algorithm on the total graph and GPS on the bipartite graph perform relative to no reordering.
\Cref{fig:reach-time} illustrates the reachability time in seconds and \Cref{fig:peak-nodes} shows the number of peak nodes.
Both scatter plots have logarithmic axes and contain data of the 110 completed models for both algorithms.
If a point lies below the line $x=y$, the result with reordering is better.
The plots also show locally weighted regression lines.
These lines indicate that for larger models the effect of reordering is more dramatic than for smaller models.
In larger cases we see an improvement in peak nodes of factors 1000 to 10000, and time of factors 10 to 100. 
There are however some models that do not benefit from the two reordering algorithms.

\section{Conclusion and Related Work}\label{sec:conclusion}
In \Cref{sec:results} we have shown that bandwidth and wavefront reduction is clearly useful
in symbolic model checking. The best algorithm for variable reordering is Sloan.
Empirical observation of results presented in this paper and the results of the 2015 model checking contest (without the traditional nodal ordering algorithms in \ltsmin) show
a big improvement and at least on par with other competitors in the statespace category.
There are two branches of related work. The first
are other model checkers, such as \smart \cite{DBLP:journals/sigmetrics/CiardoMW09},
\marcie \cite{HRS13} and \nusmv \cite{cimatti2002:nusmv2}. The \smart tool employs advanced
saturation algorithms, which can be used to confirm whether bandwidth and wavefront reduction is useful in other model checkers as well.
Readily available variable ordering algorithms are hard to find, however the \marcie model checker 
implements the Noack \cite{Liu09} variable ordering algorithm, which can be used
to compare our proposed algorithms with.

Our approach works for disjunctive partitioning schemes,
the question remains however, whether or not our method also works for conjunctive \cite{burch1991symbolic} partitioning schemes, such as in \nusmv.
Furthermore bandwidth and wavefront reduction may be applicable to SAT/SMT solving, where rows in the matrix are clauses and columns are variables.
Of interest is whether or not our proposed method can achieve similar results as the \force \cite{DBLP:conf/glvlsi/AloulMS03} heuristic, which reduces the variable cut and \emph{span}.

The second branch of related work is other bandwidth and wavefront reduction algorithms.
Kaveh \cite{eltit} discusses many different graph transformations of the adjacency graph on which nodal ordering algorithms
can be run. We picked only the total graph, because running nodal ordering
algorithms on total graphs does not require modifying these algorithms. Reid et al. \cite{DBLP:journals/siammax/ReidS06} provide
two more methods of symmetrizing an asymmetric matrix $\bm{A}$, namely $\bm{A} + \bm{A^T}$ and $\bm{A} \cdot \bm{A^T}$.
Additionally the authors provide a modified Cuthill McKee algorithm that can be run on an asymmetric matrix directly, available in the HSL library\footnote{\url{http://www.hsl.rl.ac.uk}}.
A survey \cite{mafteiu2014bandwidths} covers the state of the art in bandwidth reduction, including metaheuristic
algorithms, of which many have been developed in the past decade.

\begin{wrapfigure}{r}{.17\textwidth}%
   \vspace{-20pt}
\fbox{\includegraphics[width=.15\textwidth]{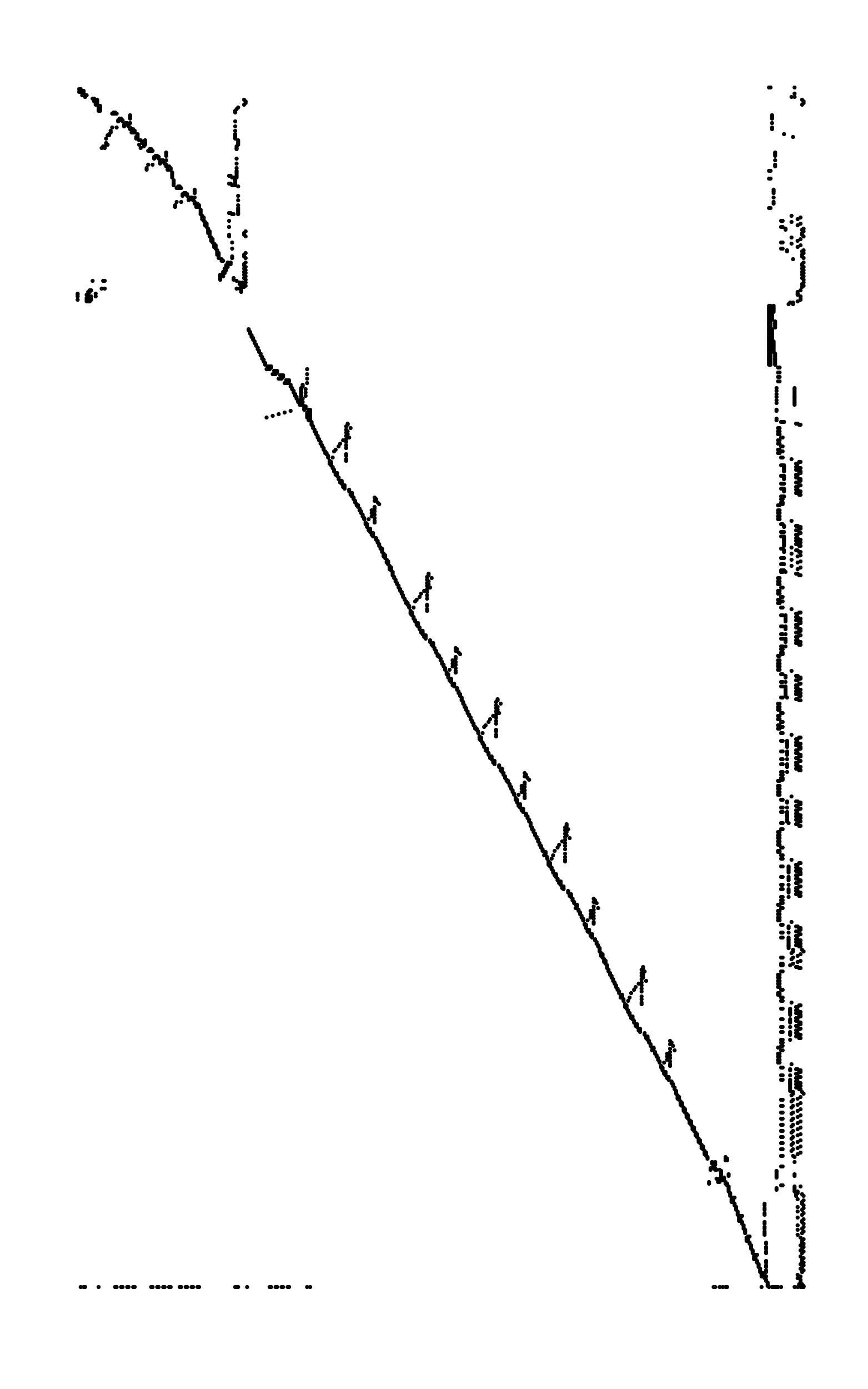}}
   \vspace{-12pt}
\caption{AMD}
\label{fig:amd}
   \vspace{-20pt}
\end{wrapfigure}%
Another nodal ordering algorithm which
is often used in iterative sparse matrix solvers is the Approximate Minimum Degree (AMD) \cite{DBLP:journals/toms/AmestoyEDD04} algorithm, 
implemented in SuiteSparse\footnote{\url{http://faculty.cse.tamu.edu/davis/suitesparse.html}}.
AMD produces matrices of the form shown in \Cref{fig:amd}. AMD can also be applied on symmetrized matrices, but
we have found AMD not applicable to symbolic model checking, judging by the form of the reordered matrices it produces.
Recently, advances have been made in parallelizing \cite{DBLP:conf/sc/KarantasisLNGP14}
nodal ordering algorithms. We think however, that dependency matrices are too small to benefit greatly from these parallelized algorithms.
\vspace{-10pt}
\subsubsection*{Acknowledgements}
We would like to thank Marcus Gerhold, Erik Kemp and Alfons Laarman for making helpful contributions to this paper.

\newpage
\bibliographystyle{splncs03}
\bibliography{main}
\end{document}